\documentclass[10pt, conference, letterpaper]{IEEEtran}

\usepackage{graphicx}
\usepackage{caption}

\usepackage{subfig}

\usepackage{blindtext}
\usepackage{adjustbox}
\usepackage{multirow}
\usepackage{color}
\usepackage{booktabs}
\usepackage{tabularx}
\usepackage{colortbl}
\usepackage{bbding}
\usepackage{tikz}
\usepackage{listings}
\usepackage{etoolbox}
\usepackage{subfig}

\usepackage{url}
\usepackage{setspace}
\usepackage[british,english]{babel}

\newcommand{\circled}[2][]{\tikz[baseline=(char.base)]
    {\node[shape = circle, draw, inner sep = 1pt]
    (char) {\phantom{\ifblank{#1}{#2}{#1}}};%
    \node at (char.center) {\makebox[0pt][c]{#2}};}}
\robustify{\circled}

\newcommand{\WS}{\texttt{WS}\xspace}
\newcommand{\HMP}{\texttt{HMP}\xspace}

\makeatletter
\def\@IEEEsectpunct{.\ \,}
\def\paragraph{\@startsection{paragraph}{4}{\z@}{1.5ex plus 1.5ex minus 0.5ex}%
{0ex}{\normalfont\normalsize\sffamily\bfseries}}

\patchcmd{\@maketitle}
  {\addvspace{0.2\baselineskip}\egroup}
  {\addvspace{-1\baselineskip}\egroup}
  {}
  {}

\makeatother

\begin{document}

\title{Energy-aware Web Browsing on Heterogeneous Mobile Platforms}

\author{
    \IEEEauthorblockN{
        Jie Ren\IEEEauthorrefmark{2},
        Ling Gao\IEEEauthorrefmark{2},
        Hai Wang\IEEEauthorrefmark{2}, and
        Zheng Wang\IEEEauthorrefmark{3}\IEEEauthorrefmark{1}
        }
    \IEEEauthorblockA{
        \IEEEauthorrefmark{2}Northwest University, China\\
        \IEEEauthorrefmark{3}MetaLab, School of Computing and Communications, Lancaster University, U.K. \\
        Emails: \IEEEauthorrefmark{2}jr@stumail.nwu.edu.cn, \IEEEauthorrefmark{2}\{gl, hwang\}@nwu.edu.cn, \IEEEauthorrefmark{3}z.wang@lancaster.ac.uk
        }
}

\IEEEoverridecommandlockouts

\maketitle
\begin{abstract}
Web browsing is an activity that billions of mobile users perform on a daily basis.  Battery life is a primary concern to many mobile users
who often find their phone has died at most inconvenient times. The heterogeneous multi-core architecture is a solution for
energy-efficient processing. However, the current mobile web browsers rely on the operating system to exploit the underlying hardware,
which has no knowledge of individual web contents and often leads to poor energy efficiency. This paper describes an automatic approach to
render mobile web workloads for performance and energy efficiency. It achieves this by developing a machine learning based approach to
predict which processor to use to run the web rendering engine and at what frequencies the processors  should operate. Our predictor learns
offline from a set of training web workloads. The built predictor is then integrated into the browser to predict the optimal processor
configuration at runtime, taking into account the web workload characteristics and the optimisation goal: whether it is load time, energy
consumption or a trade-off between them. We evaluate our approach on a representative ARM big.LITTLE mobile architecture using the hottest
500 webpages. Our approach achieves 80\% of the performance delivered by an ideal predictor.
We obtain, on average, 45\%, 63.5\% and 81\% improvement respectively for load time, energy consumption and the energy delay product, when compared to the Linux heterogeneous multi-processing scheduler.
\end{abstract}


\section{Introduction}

Web browsing is a major
activity performed by mobile users on a daily basis~\cite{mobilestat}.
However, it remains an activity of high
energy consumption~\cite{d2016energy,thiagarajan2012killed}.
Heterogeneous multi-core design, such as the ARM big.LITTLE
architecture~\cite{arm}, is a solution to energy efficient
mobile processing. Heterogeneous mobile platforms integrate multiple
processor cores on the same system, where each processor is tuned for a
certain class of workloads and optimisation goals (either performance or
energy consumption).  To unlock the potential of the heterogeneous design, software applications must adapt
to the variety of different processors and make good use of the underlying
hardware, knowing what type of processors to use and at what frequency the processor should operate. This is
because the benefits of choosing the right heterogeneous core may be
large, but mistakes can seriously hurt the user experience.

The current mobile web browser implementations rely on the operating system to exploit the heterogeneous cores. The
drawback of this is that the operating system has no knowledge of the individual web workload to be rendered by the
browser; and as a result, this often leads to poor energy efficiency, draining the battery faster than
necessary~\cite{zhu2015event}. What we would like to have is a technique that can exploit the web workload
characteristics to leverage the heterogeneous cores to meet various user requirements: whether it is load time (responsive time), energy
consumption or a trade-off between them. Given the diversity of mobile architectures, we would like to have an
automatic approach to construct optimisation strategies for any given platforms with little human involvement.

This paper presents such an approach to exploit the heterogeneous mobile platform for energy efficient web browsing. In
particular, it focuses on determining -- for a given optimisation goal -- the optimal processor configuration i.e. the
type of processor cores to use to render the webpage and at what frequencies the processor cores of the system should
operate. Rather than developing a hand-crafted approach that requires expert insight into the relative costs of
particular hardware and web contents, we develop an automatic technique that can be portable across computing
environments. We achieve this by employing machine learning to automatically build predictors based on knowledge
extracted from a set of representative, training web contents. The trained models are then used at runtime to predict
the optimal processor configuration for a given workload and an optimisation target.

Our technique is implemented as an extension for the Google Chromium browser. It is applied to the hottest 500 webpages ranked by
\texttt{www.alexa.com} and is evaluated for three distinct metrics: load time, energy consumption and energy delay
product (a trade-off between load time and energy consumption). We evaluated our technique
 on a representative big.LITTLE mobile platform. Our approach delivers significant
 improvement over a state-of-the-art web-aware scheduling mechanism~\cite{YZhu13} and the Linux Heterogeneous Multi-Processing (\HMP) scheduler for all
the three metrics.

The key contribution of this paper is a novel machine learning based predictive model that can be used to optimise web
workloads across multiple optimisation goals. Our results show that significant energy efficiency for mobile web
browsing can be achieved by making effective use of the heterogeneous mobile architecture.

%

\section{Background}

\subsection{Web Rendering Process}

\begin{figure}
\begin{center}
\includegraphics[width=0.48\textwidth]{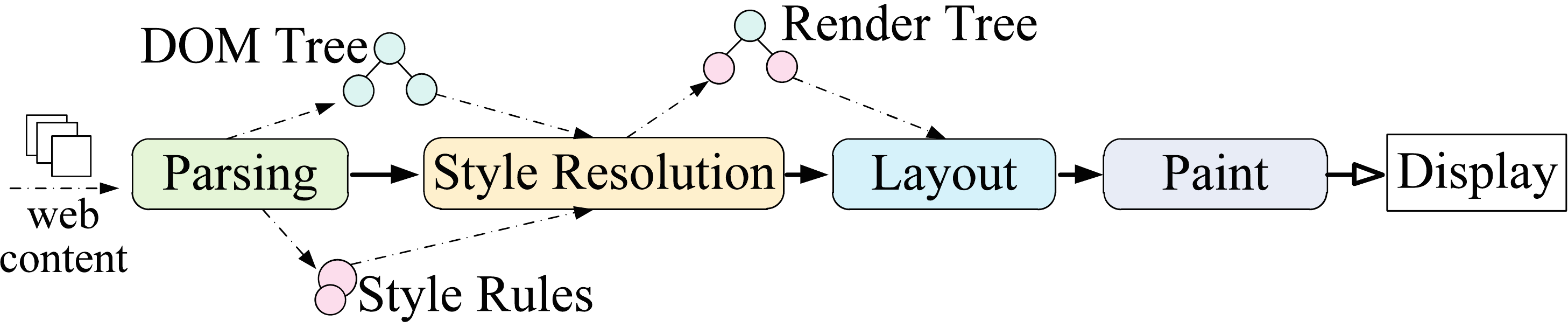}
\end{center}
\caption{The rendering process of Chromium browser.}
\vspace{-4.5mm}
\label{fig:chromium}
\end{figure}

Our prototype system is built upon Google Chromium, an open source
version of the popular Google Chrome web browser.
To render an already downloaded webpage, the Chromium rendering engine follows a number of steps:
parsing, style resolution, layout and paint.
This process is illustrated in Figure~\ref{fig:chromium}.
Firstly, the input HTML page is parsed to construct a Document Object Model (DOM) tree where
each node of the tree represents an individual HTML tag such as \texttt{<body>} or \texttt{<p>}. CSS style rules that describe how the web contents
should be presented will also be translated to the style rules. Next, the styling information and the DOM tree are
combined to build a render tree which is then used to compute the layout of
each visible element. Finally, the paint process takes in the render tree to
output the pixels to the screen.
In this work, we focus
\emph{solely} on scheduling the rendering process on heterogeneous mobile systems.

\begin{figure*}[!ht]
	\centering
	\subfloat[][Load time]{\includegraphics[width=0.23\textwidth]{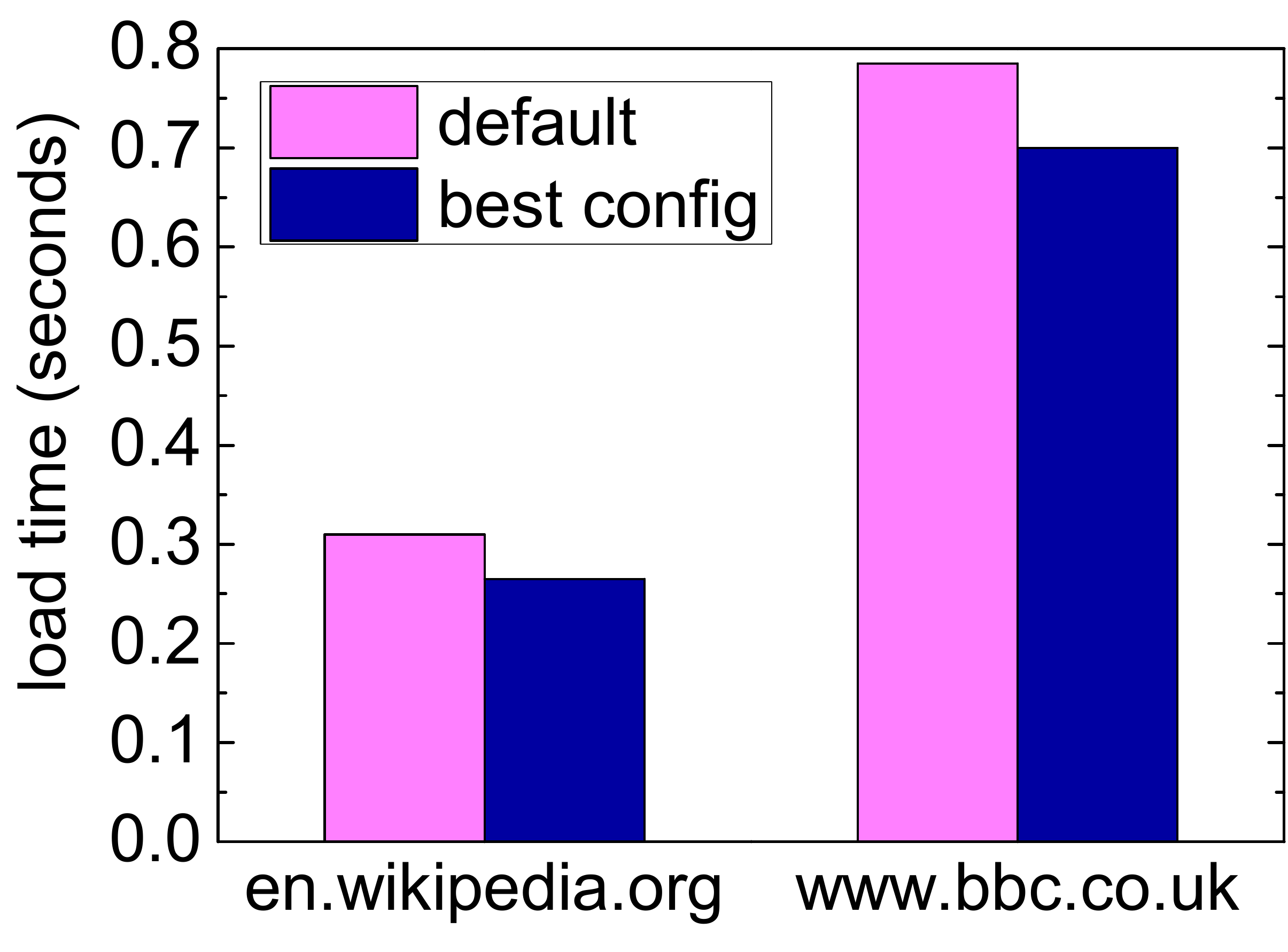}}
    \hfill
    \subfloat[][Energy consumption]{\includegraphics[width=0.23\textwidth]{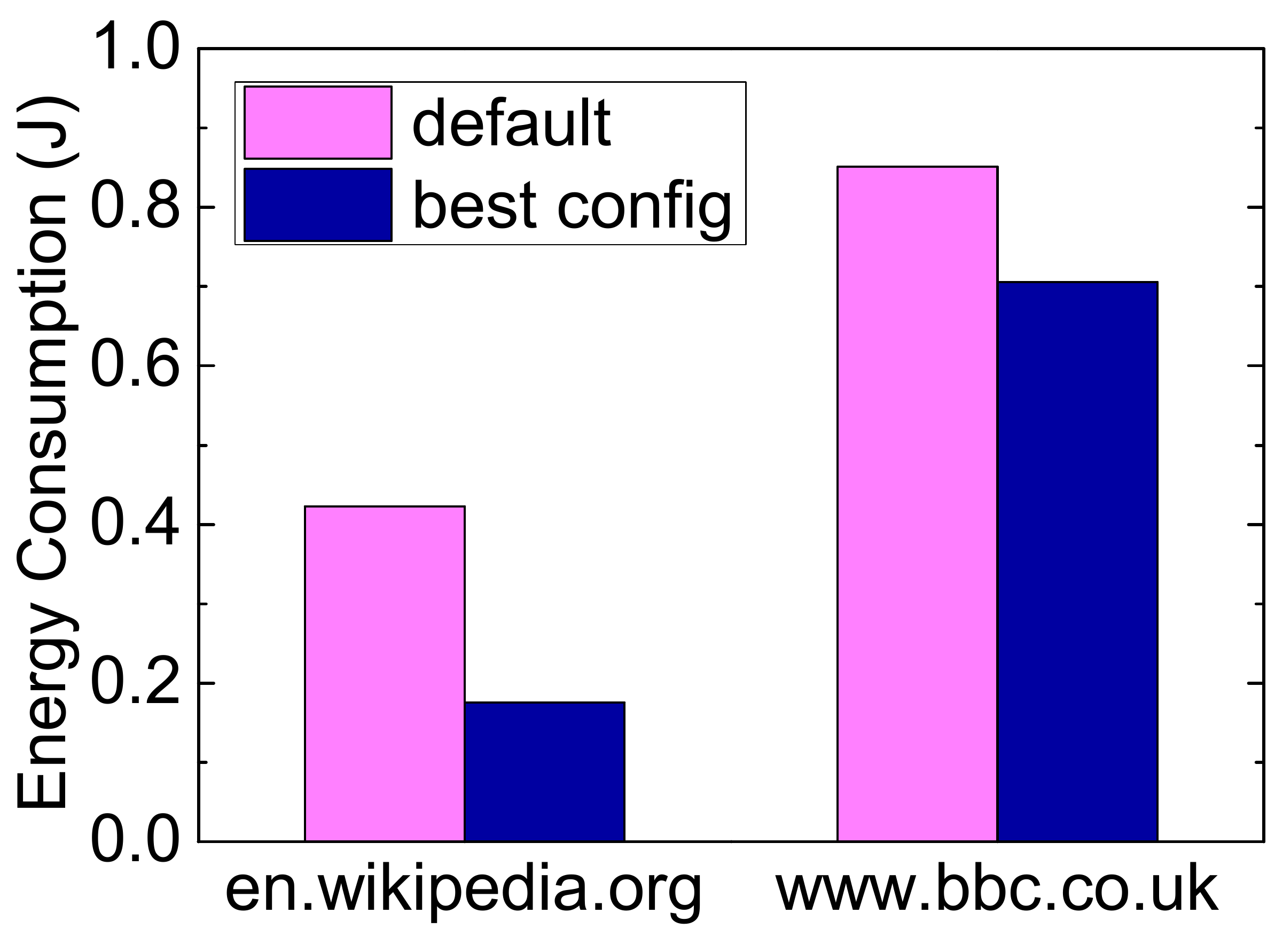}}
    \hfill
    \subfloat[][EDP]{\includegraphics[width=0.23\textwidth]{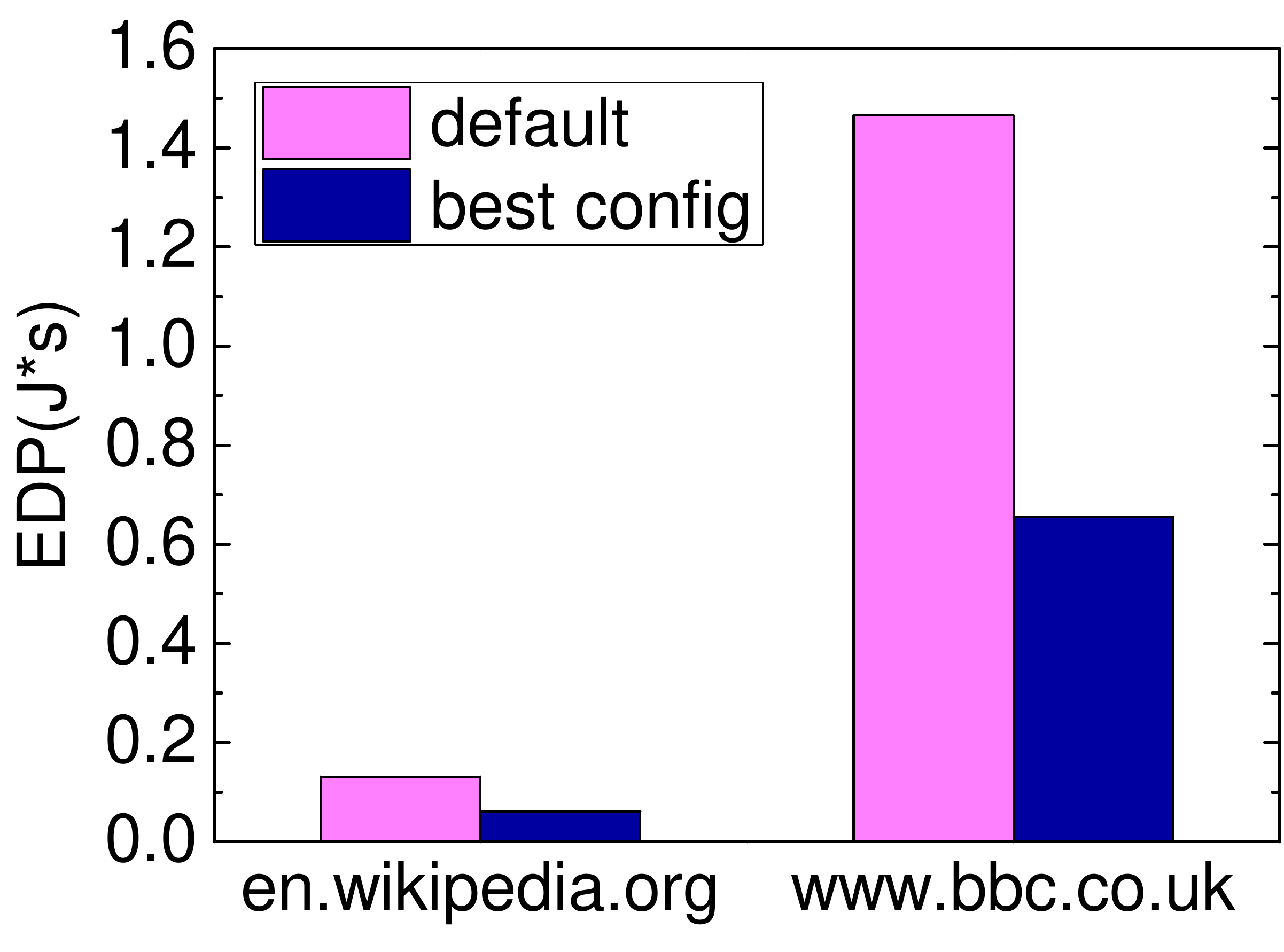}}
    \hfill
    \subfloat[][wikipedia best on bbc.co.uk]{\includegraphics[width=0.23\textwidth]{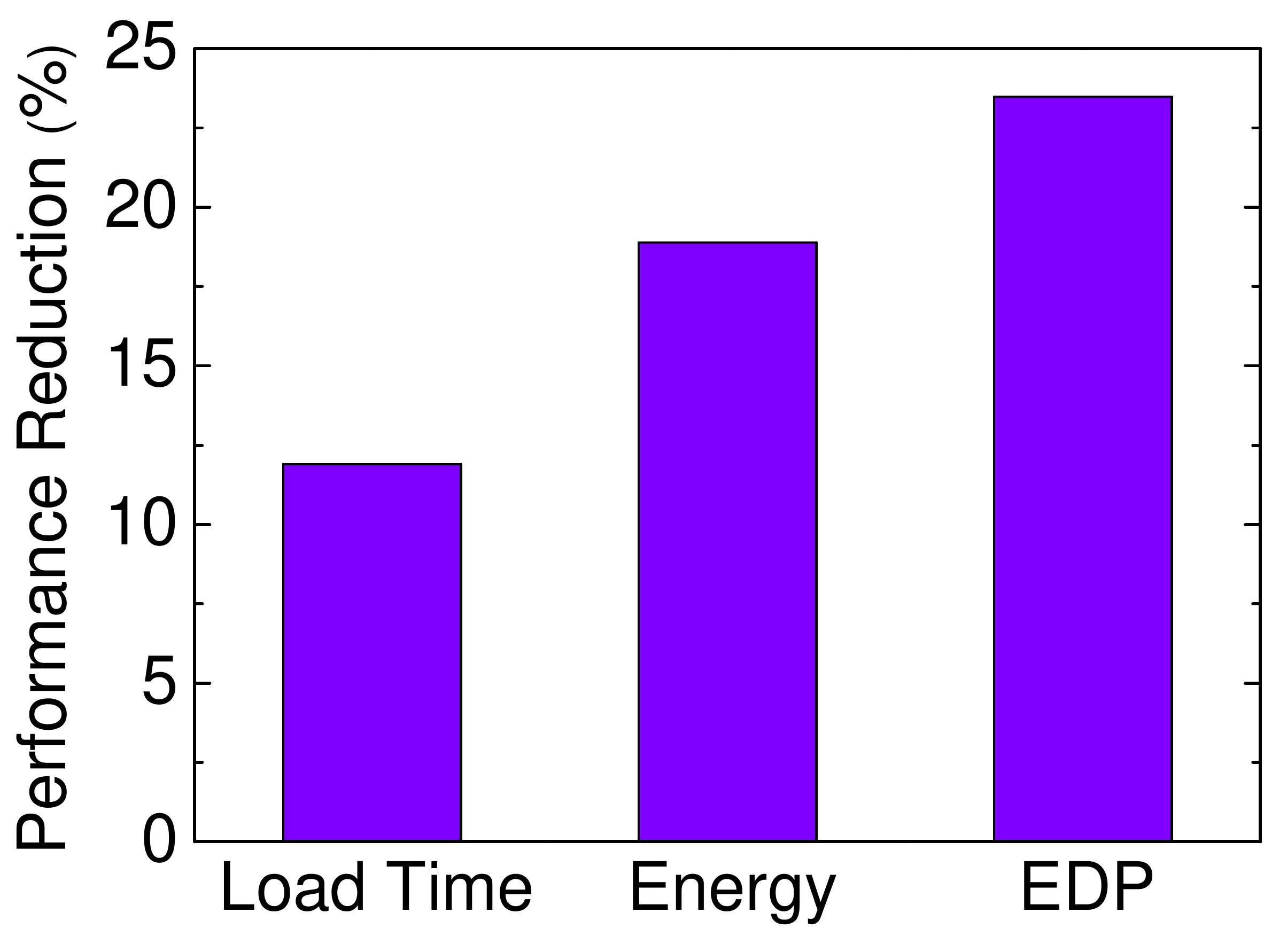}}
    \caption{Best load time (a), energy consumption (b) and EDP (c) for rendering \texttt{wikipedia} and \texttt{bbc} over to the \HMP scheduler;
    and the performance of using \texttt{wikipedia} best configurations w.r.t to the best available performance of \texttt{bbc} (d).}
    \vspace{-3mm}
    \label{fig:motivation}
\end{figure*}

\subsection{Motivation Example \label{sec:motivation}}

Consider rendering the landing page of \texttt{en.wikipedia.org} and
\texttt{www.bbc.co.uk} on an ARM big.LITTLE mobile platform. The system has a Cortex-A15 (big)
and a Cortex-A7 (little) processors, running with the Ubuntu Linux operating system (OS) (see also Section~\ref{sec:hs_setup}).
Here, we schedule the
Chromium rendering process to run on the big or little core
under various clock frequencies. We then record the best processor configuration found
for each webpage. To isolate network and disk overhead, we
have pre-downloaded and stored the webpages in the RAM and disabled the browser's cache.

Figure~\ref{fig:motivation} compares the best configuration against the Linux \HMP scheduler  for three \emph{lower is better} metrics: (a)
load time, (b) energy consumption and (c) the energy delay product (EDP), calculated as energy $\times$ load time.
Table~\ref{tab:bestConfig} lists the best configuration for each metric.
For load time, the best configuration gives 14\% and 10\% reduction for
\texttt{wikipedia} and \texttt{bbc} respectively over
the \HMP. For energy consumption, using the right processor configuration gives
a reduction of 58\% and 17\%  for \texttt{wikipedia} and \texttt{bbc}
respectively. For EDP, the best configuration
gives a reduction of over 55\%  for both websites. Clearly, there is significant room for improvement over the
OS scheduler and the best processor configuration could change from one
metric to the other.

Figure~\ref{fig:motivation} (d) normalises the best available performance of
\texttt{bbc} to the performance achieved by using the best configuration
found for \texttt{wikipedia} for each metric. It shows that the best
processor configuration could also vary across webpages. The
optimal configuration for \texttt{wikipedia} fails to deliver the best
available performance for \texttt{bbc}. In fact, there is a reduction of
11.9\%, 18.9\% and 23.5\% on load time, energy and EDP available respectively
for \texttt{bbc} when compared to using the \texttt{wikipedia-best}
configuration. Therefore, simply applying one optimal configuration found for
one webpage to another is likely to miss significant optimisation
opportunities.

This example demonstrates that using the right processor setting has a
significant impact on web browsing experience, and the optimal configuration
depends on the optimisation objective and the workload. What we need is a
technique that automatically determines the best configuration for any
webpage and optimisation goal. In the remainder of this paper, we describe
such an approach based on machine learning.

\begin{table}[!t]
\caption{Optimal processor configurations for web rendering}
\scriptsize
\begin{center}
        \begin{tabular}{lcccccc}
        \toprule
        \rowcolor[gray]{.92}& \multicolumn{2}{c}{Load time} &\multicolumn{2}{c}{Energy}& \multicolumn{2}{c}{EDP} \\
        & A15 & A7 & A15 & A7 & A15 & A7\\
        \midrule
        \texttt{en.wikipedia.org} - GHz&1.8&1.4&0.9&0.4&1.3&0.5\\
        \texttt{www.bbc.co.uk} - GHz&1.6&1.4&1.0&0.3&1.5&0.4\\
          rendering engine &\Checkmark& &\Checkmark& &\Checkmark& \\
        \bottomrule
        \end{tabular}
\end{center}
\label{tab:bestConfig}
\vspace{-5mm}
\end{table}

\section{Overview of our approach}

\begin{figure*}
\begin{center}
\includegraphics[width=0.76\textwidth]{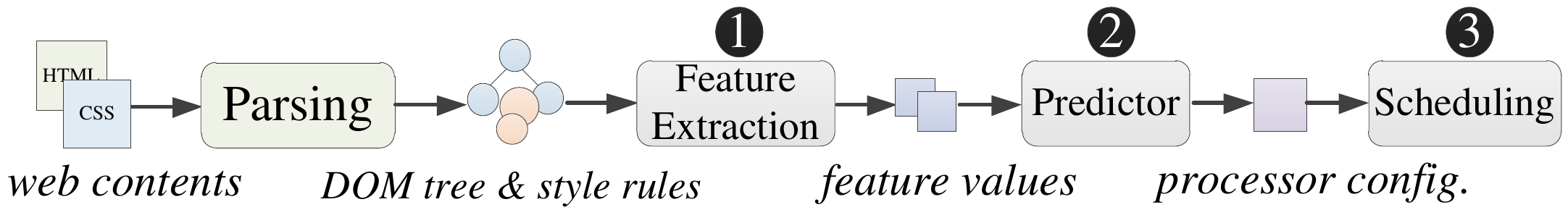}
\end{center}
\vspace{-2mm}
\caption{Our three-stage approach for predicting the best processor configuration and scheduling the rendering process. }
\vspace{-2.5mm}
\label{fig:workflow}
\end{figure*}


Figure 3 depicts our three-stage approach for predicting the right processor
configuration when rendering a webpage.  After the web contents (e.g. the HTML source,
CSS files and javascripts) are downloaded, they will be  parsed to construct a
DOM tree together with style rules. This is performed by the default parser of the browser. Our approach begins from extracting information (\emph{termed as feature extraction}),
from the DOM tree and style data to characterise the web workload.
This information (or features) includes counting different HTML tags, DOM nodes and style rules. A complete list of the features is given in Table~\ref{tab:selected_features}.
Next, a machine learning based predictor
(that is built off-line) takes in these feature values and predicts which core to use to
run the rendering engine and
at what frequencies the processors of the platform should operate.
Finally, we configure
the processors and schedule
the rendering engine to run on the predicted core.

Our approach is implemented as a web browser extension which will be invoked as soon as a DOM tree is constructed.
Re-prediction and rescheduling will be triggered if there are significant changes
of the DOM tree structure, so that we can adapt to the change of web contents.
Note that we let the operating system to schedule other web browser threads such as the input/output process.

\paragraph*{Optimisation Goals} In this work we target three important optimisation metrics:
(a) load time (which aims to
render the webpage as quick as possible), (b) energy consumption (which aims to
use as less energy as possible) and (c) EDP (which aims to balance the load time and energy consumption).
For each metric, we construct a predictor using the same learning methodology described in the next section.


%
%

\section{Predictive Modeling}
Our model for predicting the best processor configuration is a
Support Vector Machine (SVM) classifier using a radial basis function kernel~\cite{vapnik1998statistical}. We
have evaluated a number of alternative modeling techniques, including
regression, Markov chains, K-Nearest neighbour, decision trees, and artificial neural networks. We chose
SVM because it gives the best performance, can model both linear and
non-linear problems and the model produced by the learning algorithm is deterministic. The input to
our model is a set of features extracted from the DOM tree and style rules.
The output of our model is a label that indicates the optimal core to use to run the rendering engine
and the clock frequencies of the CPU cores of the system.

Building and using such a model follows
the well-known 3-step process for supervised machine learning: (i) generate training data (ii) train a predictive
model (iii) use the predictor, described as follows.

\subsection{Training the Predictor}
\begin{figure}[t!]
  \centering
  \includegraphics[width=0.45\textwidth]{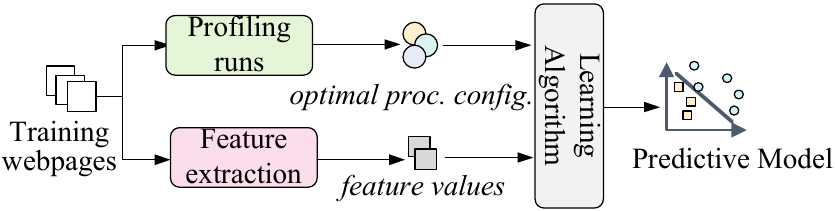}\\

  \caption{Training the predictor.}\label{fig:training}
  \vspace{-1mm}
\end{figure}

\begin{table}[!t]
\caption{Useful processor configurations per metric.}
\vspace{-2mm}
\small
\begin{center}
        \begin{tabular}{cccccc}
        \toprule
        \rowcolor[gray]{.92} \multicolumn{2}{c}{Load time}&\multicolumn{2}{c}{Energy} &\multicolumn{2}{c}{EDP}\\
         A15 & A7 & A15 & A7 & A15 & A7\\
        \midrule
         \rowcolor[gray]{.92}\textbf{1.6}&1.4&\textbf{0.8}&0.4&\textbf{1.3}&0.4\\
        \textbf{1.7}&1.4&\textbf{0.9}&0.4&\textbf{1.4}&0.4\\
        \rowcolor[gray]{.92}\textbf{1.8}&1.4&\textbf{1.0}&0.4&\textbf{1.5}&0.4\\
        \textbf{1.9}&1.4&0.3&\textbf{1.1}&0.5&\textbf{1.2}\\
        \rowcolor[gray]{.92}-&-&0.3&\textbf{1.2}&0.5&\textbf{1.3}\\
        -&-&0.3&\textbf{1.3}&0.5&\textbf{1.4}\\
        \bottomrule
        \end{tabular}
\end{center}
\vspace{-5mm}
\label{tab:trainingConfig}
\end{table}

\begin{table}[t!]
\caption{Raw web features}
\small
\centering
        \begin{tabular}{rll}
        \toprule
        \multirow{2}{*}{DOM Tree} & \#DOM nodes & depth of tree \\
                & \#each HTML tag & \#each HTML attr. \\
        \rowcolor[gray]{.92}  & \#rules  & \#each property \\
        \rowcolor[gray]{.92}  \multirow{-2}{*}{Style Rules} & \#each selector pattern & \\
        Other  &  webpage size (KB) & \\
        \bottomrule
        \end{tabular}
\label{tab:feature}
\vspace{-5mm}
\end{table}

Figure~\ref{fig:training} depicts the process of using training webpages to
build a SVM classifier for one of the three optimisation metrics. Training
involves finding the best processor configuration and extracting feature values for
each training webpage, and learning a model from the training data.

\paragraph*{Generate Training Data}
We use over 400 webpages to train a SVM model. These webpages are the landing
page of the top 500 hottest websites ranked by \texttt{alexa}~\cite{alexa}.
These websites cover a wide variety of areas, including shopping,
video, social network, search engine, E-commerce, news etc. Whenever
possible, we used the mobile version of the website. Before training, we have
pre-downloaded the webpages from the Internet and stored the content in a RAM
disk. For each webpage, we exhaustively execute the rendering engine with
different processor settings and record the best performing configuration for
each optimisation metric. We then label each best-performing
configuration with a unique number. Table~\ref{tab:trainingConfig} lists the
processor configurations found to be useful on our hardware platform. For
each webpage, we also extract the values of a selected set of features (described
in Section~\ref{sec:web_features}).

\paragraph*{Building The Model} The feature
values together with the labelled processor configuration are supplied
to a supervised learning algorithm. The learning algorithm tries to find a
correlation from the feature values to the optimal configuration and outputs
a SVM model. Because we target three optimisation metrics in this paper,
we have constructed three SVM models -- one for each optimisation metric.
Since training is only performed once at the factory, it is a
\emph{one-off} cost. In our case the overall training process takes less than
a week using two identical hardware platforms.

One of the key aspects in building a successful predictor is
finding the right features to characterise the input data.
This is described in the next section which is followed by sections  describing how to use the predictor at runtime.

\subsection{Web Features \label{sec:web_features}}

\begin{table}[!t]
\caption{Selected web features}
\small
\centering
        \begin{tabular}{lp{6cm}}
        \toprule
        \#HTML tag & a, b, br, button, div, h1, h2, h3, h4, i, iframe, li, link, meta, nav, img,
        noscript, p, script, section, span, style, table, tbody\\
        \rowcolor[gray]{.92} \#HTML attr & alt, async, border, charset, class, height, content, href, media, method, onclick, placeholder, property, rel, role, style, target, type, value, background, cellspacing, width, xmlns, src\\
        \#Style selector & class, descendant, element, id\\
        \rowcolor[gray]{.92} \#Style rules &  background.attachment/clip/color/image, background.repeat.x/y, background.size,
        background.border.image.repeat/slice/source/width, font.family/size/weight, color, display, float\\
        Other info. & DOM tree depth,  \#DOM nodes, \#style rules,  size of the webpage (Kilobytes)\\
        \bottomrule
        \end{tabular}
\label{tab:selected_features}
\vspace{-2mm}
\end{table}

Our predictor is based on a number of features extracted from the HTML and
CSS attributes. We started from 948 raw features that can be collected
at runtime from Chromium. Table~\ref{tab:feature} lists the raw features
considered in this work. These are chosen based on our intuitions of what factors can affect scheduling.
For examples, the DOM tree structures (e.g. the number of nodes, depth of
the tree, and HTML tags) determine the complexity and layout of the
webpage; the style rules determine how elements (e.g. tables and fonts) of
the webpage should be rendered; and the larger size of the webpage the longer
the rendering time is likely to be.

\paragraph*{Feature Selection}
To build an accurate predictor using supervised learning, the training sample size typically needs to be
at least one order of magnitude greater than the number of features. Given the size of our training
examples (less than 500 webpages), we would like to reduce the number of
features to use.  We achieve this by removing features that carry little or
redundant information. For instances, we have removed features of HTML tags or attributes that
are found to have little impact on the rendering time or processor selections.
Examples of those tags are \texttt{<def>}, \texttt{<em>} and \texttt{<body>}.
We have also constructed a correlation coefficient matrix to quantify the
correlation among features to remove similar features. The correlation
coefficient takes a value between $-1$ and $1$, the closer the coefficient is to
$+/-1$, the stronger the correlation between the features. We
removed features that have a correlation coefficient greater than 0.75 (ignore
the sign) to any of the already chosen features.  Exemplary similar features
include the CSS styles \texttt{<marginTop>} and \texttt{<marginRight>} which
often appear as pairs. Our feature selection process results in 73 features listed in Table~\ref{tab:selected_features}.

\paragraph*{Feature Extraction}
To extract features from the DOM tree, our extension first obtains a reference
for each DOM element by traversing the DOM tree and then uses the Chromium
API, \texttt{document.getElementsByID}, to collect node information. To
gather CSS style features, it uses the \texttt{document.styleSheets} API to
extract CSS rules, including selector and declaration objects.

\paragraph*{Feature Normalisation}
Before feeding the feature values to the learning algorithm, we scale
the value of each feature to the range of 0 and 1. We also record the \emph{min} and \emph{max}
values used for scaling, which then can be used to normalise feature values extracted from the
\emph{new} webpage during runtime deployment (described in the next sub-section).

\subsection{Runtime Deployment}

%

\begin{figure}
\begin{center}
\includegraphics[width=0.48\textwidth]{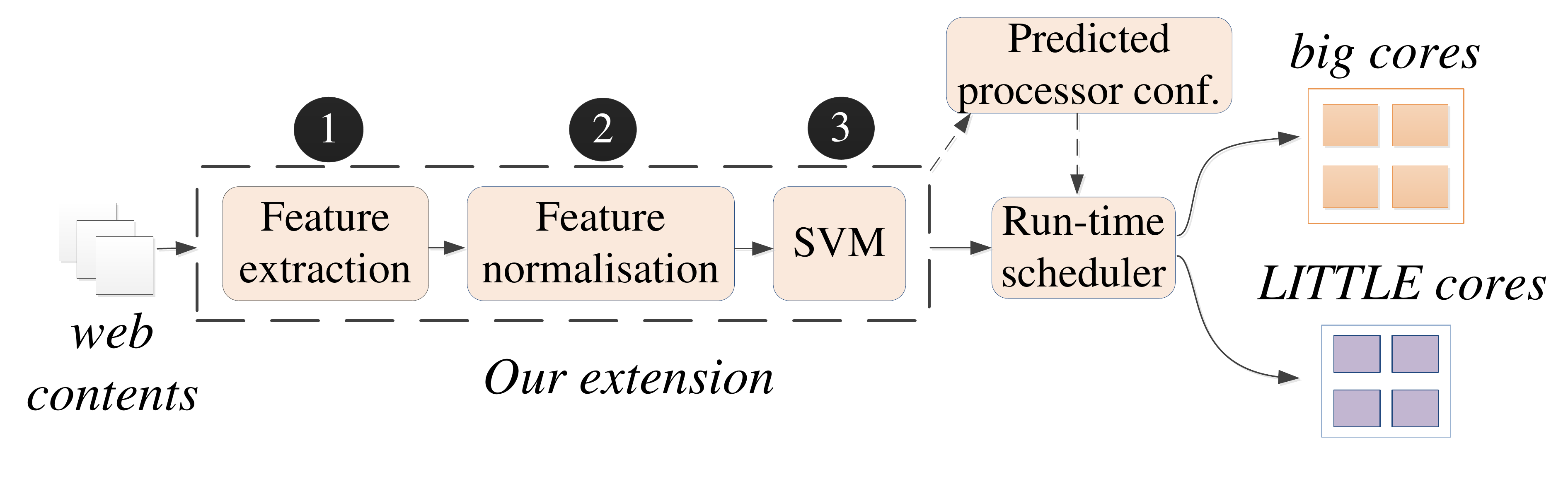}
\end{center}
\vspace{-3mm}
\caption{Runtime prediction and processor configuration.
}

\label{fig:deployment}
\vspace{-3mm}
\end{figure}



Once we have built the models as described above, we can use them to predict
the best processor configuration for any \emph{new}, \emph{unseen} web
contents. The prediction is communicated to a scheduler running as an OS service to
move the rendering process to the predicted core and set the processors to
the predicted frequencies.

Figure~\ref{fig:deployment} illustrates the process of runtime prediction and
task scheduling. During the parsing stage, which takes less than 1\% of the
total rendering time~\cite{meyerovich2010fast}, our extension firstly
extracts and normalises the feature values, and uses a SVM classifier to
predict the optimal processor configuration for a given optimisation goal.
The prediction is then passed to the runtime scheduler to perform task
scheduling and hardware configuration. The overhead of extracting features,
prediction and configuring processor frequency is small. It is less than 20
ms which is included in our experimental results.

As the DOM tree is constructed incrementally by the parser, it can change throughout the duration of rendering.
To make sure our approach can adapt to the change of available information, re-prediction and rescheduling will be triggered if the DOM tree
is significantly different from the one used for the last prediction. The difference is calculated by counting the number of DOM nodes between the currently used tree and the newly available one.
 If
the difference is greater than 30\%, we will make a new prediction
using feature values extracted from the new DOM tree and style rules. We
have observed that our initial prediction often remains unchanged, so
rescheduling and reconfiguration rarely happen in our experiments.

\subsection{Example}

\begin{table}[!t]
\caption{None-zero feature values for \texttt{en.wikipedia.org}.}
\small
\begin{center}
        \begin{tabular}{rrr}
        \toprule
        \textbf{Feature} & \textbf{Raw value} & \textbf{Normalised value}\\
        \midrule

       \rowcolor[gray]{.92} \#DOM nodes & 754 &0.084\\
        depth of tree & 13 &0.285\\
       \rowcolor[gray]{.92} number of rules & 645 &0.063\\
        web page size &2448&0.091\\

        \rowcolor[gray]{.92} \#div & 131 & 0.026\\
        \#h4 & 28 & 0.067\\
       \rowcolor[gray]{.92} \#li & 52 & 0.031\\
        \#link & 10 & 0.040\\
        \rowcolor[gray]{.92} \#script & 3 & 0.015\\

        \#href & 148 & 0.074\\
        \rowcolor[gray]{.92} \#src & 36 & 0.053\\

        \#background.attachment & 147 & 0.040\\
        \rowcolor[gray]{.92} \#background.color & 218 & 0.058\\
        \#background.image & 148 & 0.039\\

        \rowcolor[gray]{.92} \#class & 995 & 0.045\\
        \#descendant & 4454 & 0.0168\\
        \rowcolor[gray]{.92} \#element & 609 & 0.134\\
        \#id & 4 & 0.007\\
        \bottomrule
        \end{tabular}
\end{center}
\vspace{-5mm}
\label{tab:normalized}
\end{table}

As an example, consider rendering the landing page of \texttt{wikipedia}
for energy consumption. This scenario is most useful
when the mobile phone battery is low but the user still wants to retrieve information from \texttt{wikipedia}.  For this example, we have constructed a SVM model
for energy using ``cross-validation" (see Section~\ref{sec:evluation_method}) by
excluding the webpage from the training example.

To determine the optimal processor configuration for energy consumption, we first extract values of the 73 features
listed in Table~\ref{tab:selected_features} from the DOM tree and CSS style objects. The feature values will then be
normalised as described in Section~\ref{sec:web_features}. Table~\ref{tab:normalized} lists some of the non-zero values
for this website, before and after normalisation. These feature values will be fed into the offline-trained SVM model
which output a labeled processor configuration ($<$A15, 0.9, 0.4$>$ in this case) indicating the optimal configuration
is running the rendering process on the big core at 900 MHz and the little core should operate at the lowest possible
frequency 400 MHz. This prediction is indeed the optimal configuration (see also Section~\ref {sec:motivation}).
Finally, the processor configuration is communicated to a runtime scheduler to configure the hardware platform. For
this example, our approach is able to reduce 58\% of the
energy consumption when comparing to the Linux \HMP scheduler.

\begin{figure}[!t]
	\centering
    \subfloat[][\#DOM nodes]{\includegraphics[width=0.24\textwidth]{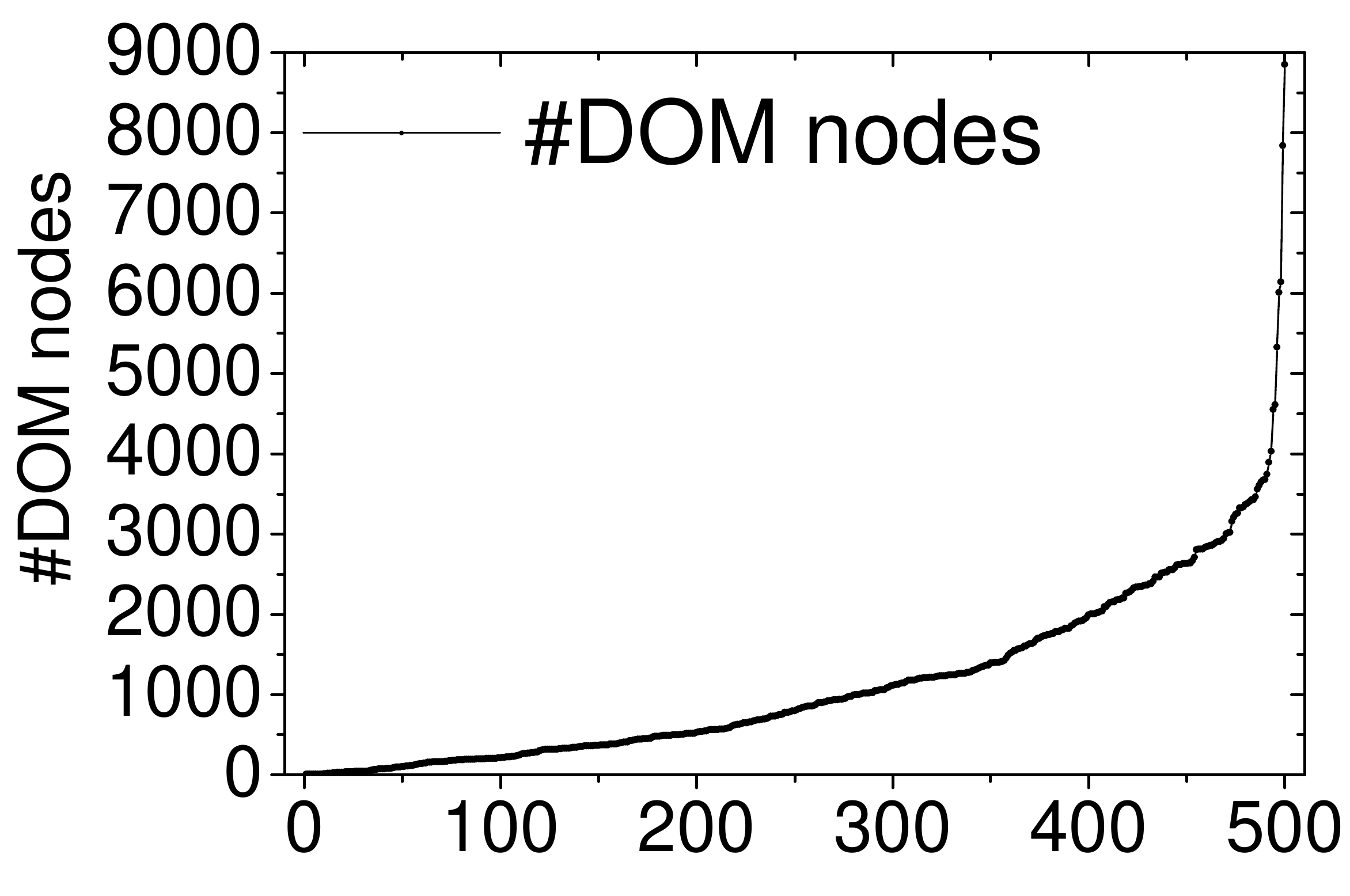}}
    \hfill
    \subfloat[][Webpage size]{\includegraphics[width=0.24\textwidth]{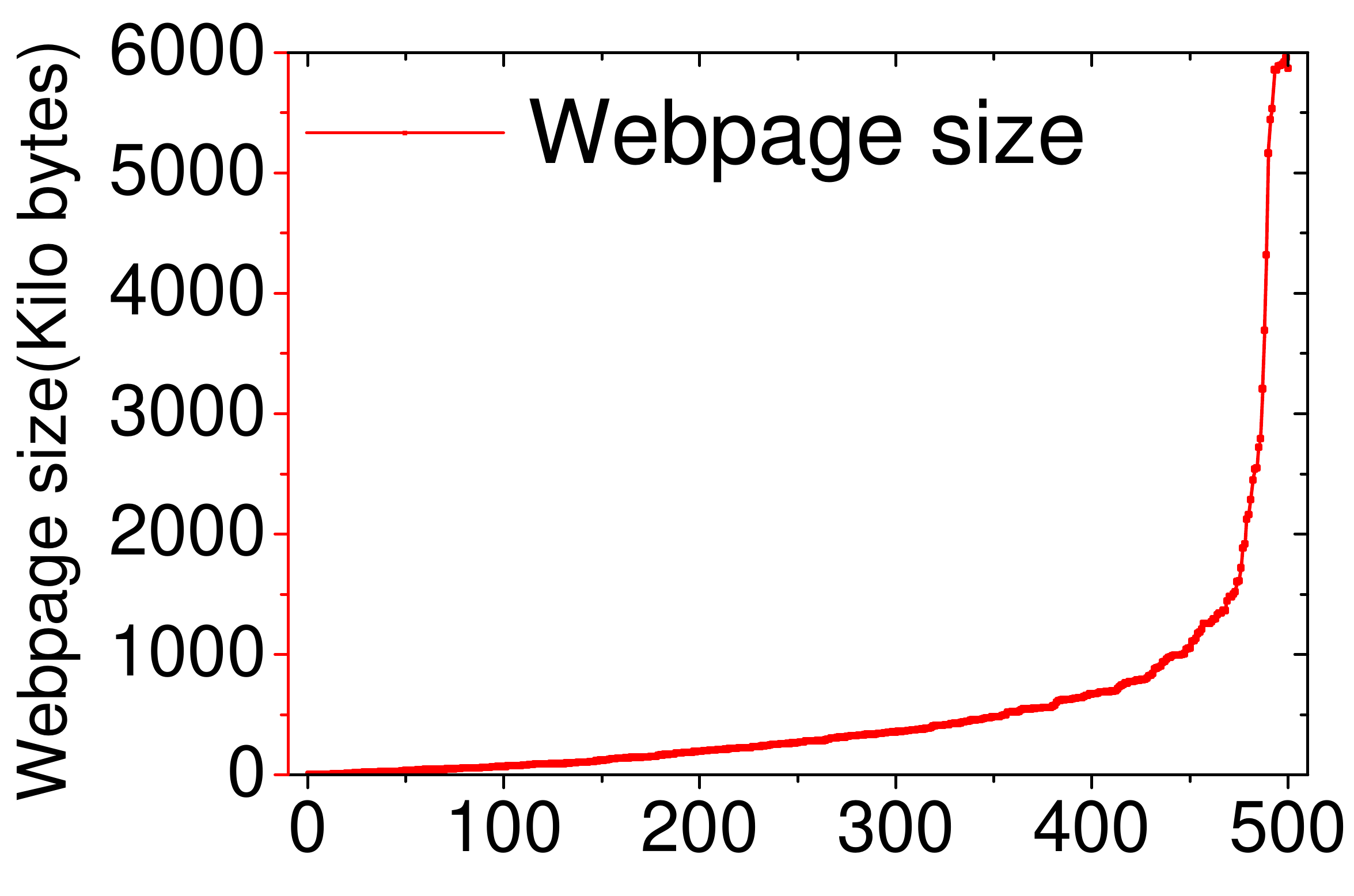}}
    \caption{The \#DOM nodes (a) and webpage size (b) for the webpages used in the experiments.}
    \vspace{-3mm}
    \label{fig:diversity}
\end{figure}

\begin{figure*}[t]
	\centering
    \subfloat[Load time]
    {\includegraphics[width=0.33\textwidth]{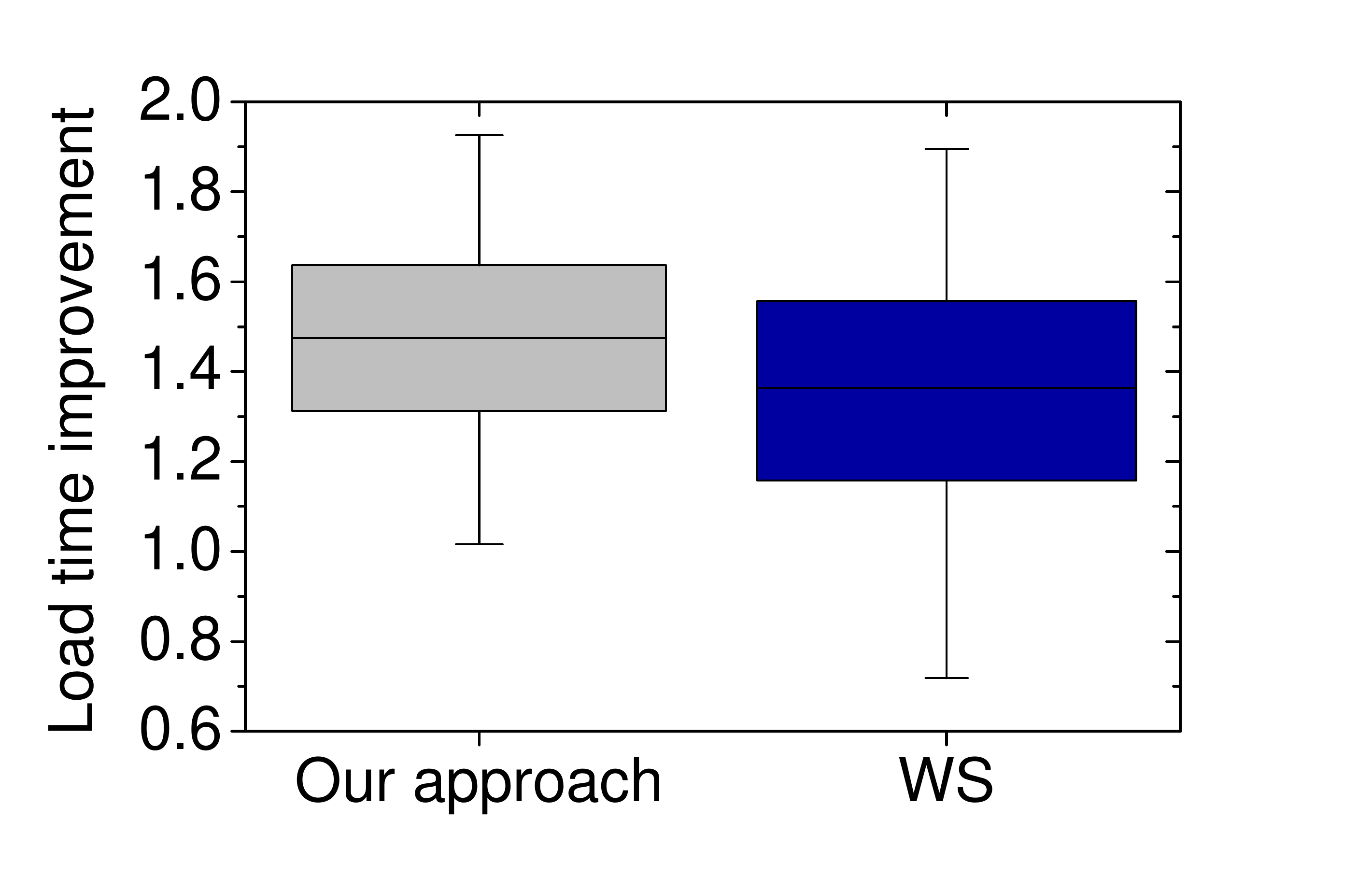}}
    \hfill
	\subfloat[Energy reduction]
    {\includegraphics[width=0.33\textwidth]{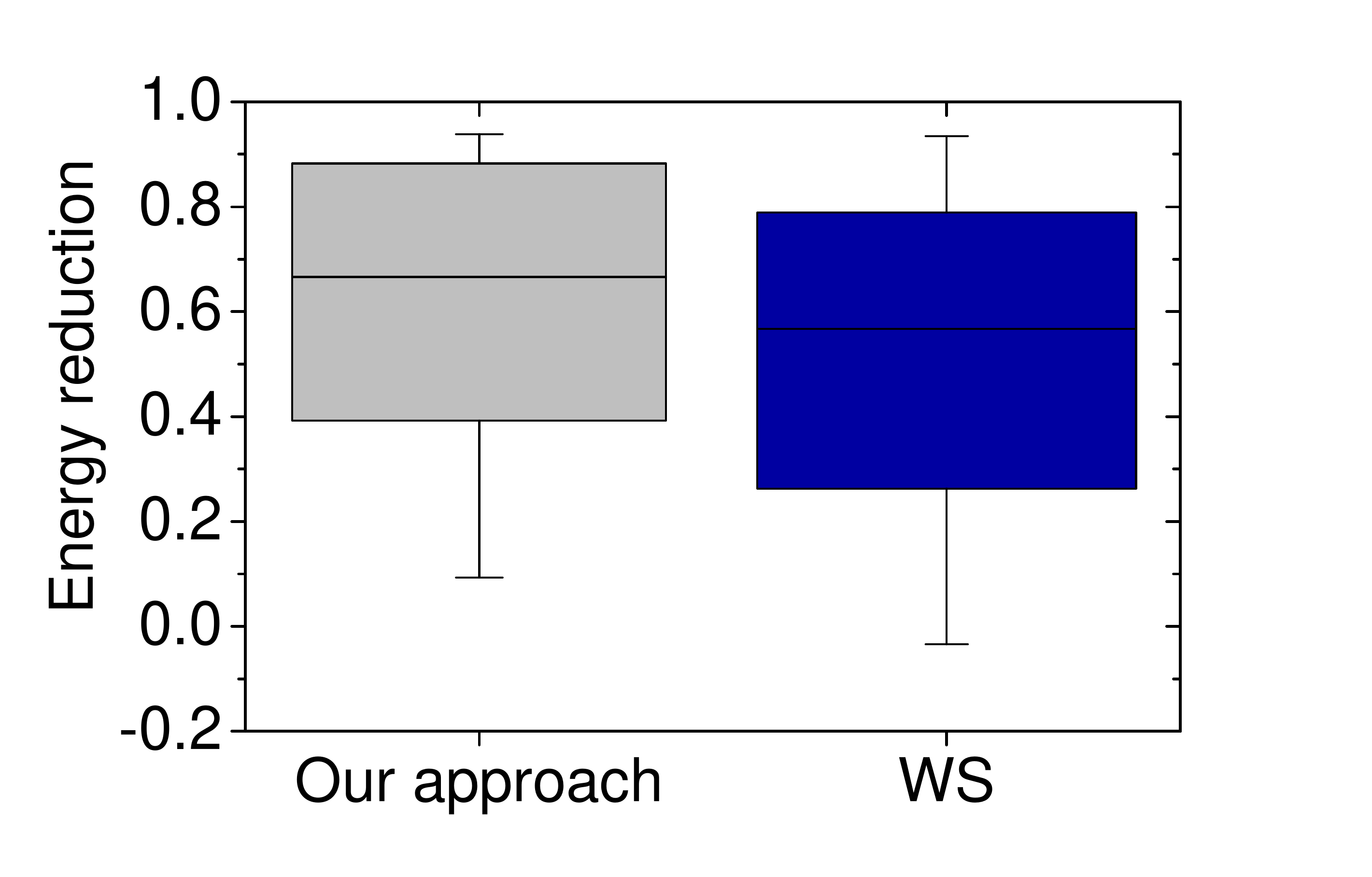}}
    \hfill
    \subfloat[EDP]
    {\includegraphics[width=0.33\textwidth]{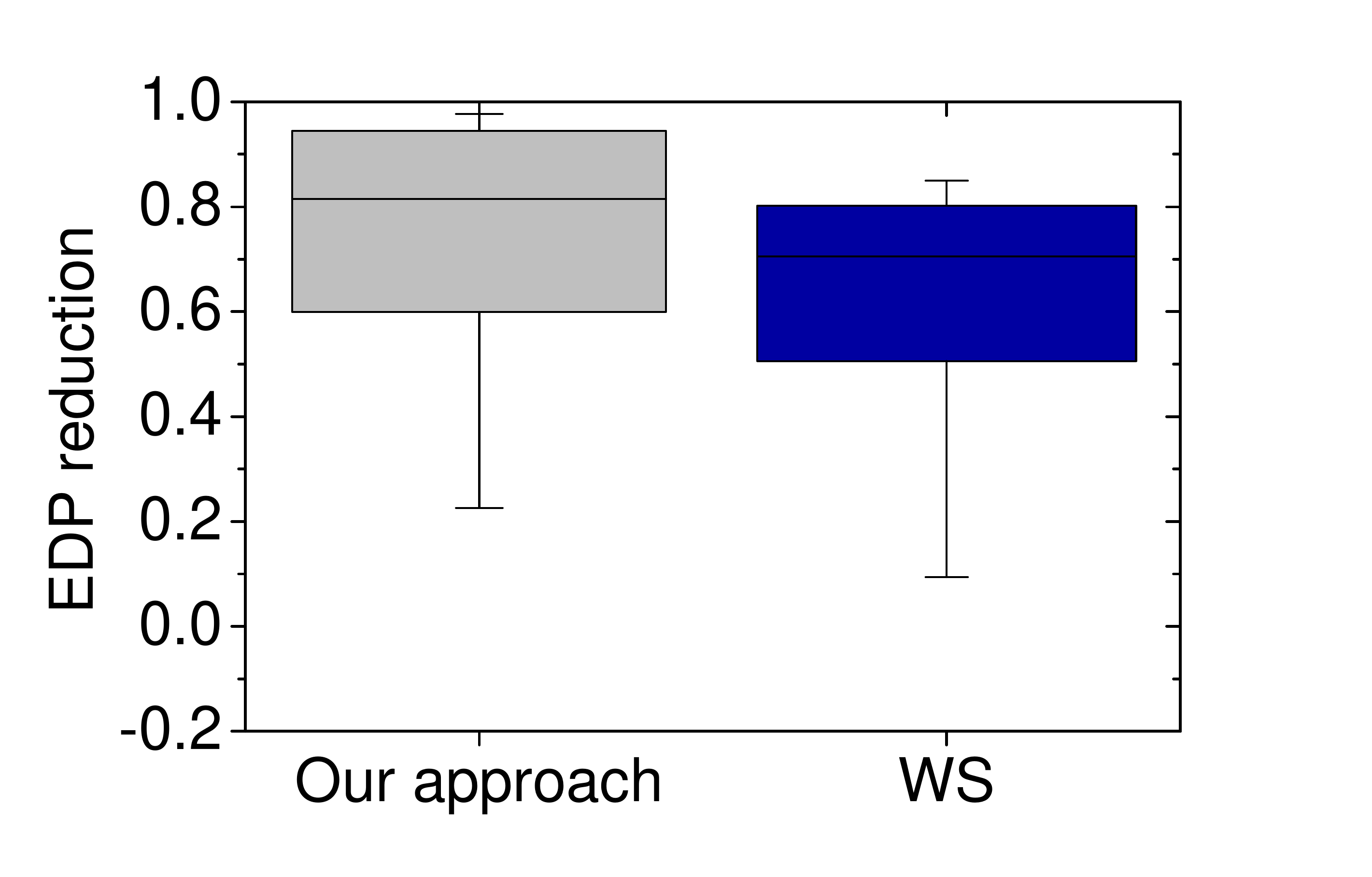}}
    \hfill

    \caption{Achieved performance for load time (a), energy consumption (b) and EDP (c) over the Linux \HMP scheduler.}
    \label{fig:overall}
    \vspace{-4mm}
\end{figure*}

\section{Experimental Setup \label{sec:setup}}

\subsection{Hardware and Software \label{sec:hs_setup}}
\paragraph*{Evaluation System} Our hardware evaluation platform is an Odroid XU3
mobile development board with an A15 (big) processor and an A7 (little) processor. The board has 2
GB LPDDR3 RAM and 64 GB eMMC storage.
Table~\ref{tbl:config} gives detailed
information of the hardware platform.
We chose this platform
as it is a representative big.LITTLE architecture implementation.
For example, the Samsung Galaxy S4 phone uses the same architecture.
The board runs the Ubuntu 14.04 LTS Linux OS.
We implemented our model
as an extension to Chromium (version 48.0) which was compiled using the gcc compiler (version 4.6).
As the
current implementation of Google Chromium for Android does not support
extensions, we did not evaluate our approach on the Android OS.

\paragraph*{Webpages} We used the landing page of the top 500 hottest websites ranked by \texttt{www.alexa.com}. Whenever possible, we used
the mobile version of the website for evaluation. To isolate network and disk overhead, we have downloaded and stored the
webpages in a RAM disk. We also disabled the browser's cache in the experiments.
Figure~\ref{fig:diversity} shows the number of DOM nodes and the size for the 500 webpages used in our evaluation. As can be seen from this diagram, the webpages range from small (4 DOM nodes and 40 Kilobytes) to large (over 8,000 DOM nodes and over 5 MB).


\begin{table}[t!]
\begin{center}
\caption{Hardware platform}
\small
\label{tbl:config}
\begin{tabular}{llll}
\toprule
&big CPU & LITTLE CPU & GPU \\
\midrule
\textbf{Model} & Cortex-A15 & Cortex-A7  & Mali-T628 \\
\textbf{Core Clock} & 2.0 GHz & 1.4 GHz & 533 MHz \\
\textbf{Fore Count} & 4 & 4 & 8 \\
\bottomrule
\end{tabular}
\end{center}
\vspace{-6mm}
\end{table}
\subsection{Evaluation Methodology \label{sec:evluation_method}}

\paragraph*{Predictive Modelling Evaluation}
We use \emph{leave-one-out} cross-validation to evaluate
our machine learning model.
This means we remove the target webpage to be predicted
from the training example set and then build a model
based on the remaining webpages. We repeat this procedure
for each webpage in turn. It is a standard evaluation
methodology, providing an estimate of the generalisation ability
of a machine-learning model in predicting \emph{unseen}
data.
\vspace{-1mm}
\paragraph*{Comparisons} We compare our approach to two alternative
approaches, a state-of-the-art web-aware scheduling mechanism~\cite{YZhu13}
(\emph{referred as \WS  hereafter}), and the Linux Heterogeneous Multi-Processing (\HMP)
scheduler which is designed for the big.LITTLE architecture to enable the use of
all different CPU cores at the same time. \WS uses a regression model built from
the training examples to estimate webpage load time and energy consumption
under different processor configurations. The model is used to find the best
configuration by enumerating all possible configurations.
\vspace{-1mm}
\paragraph*{Performance Report}
We profiled each webpage under a processor configuration multiple times
and report the \emph{geometric mean} of each evaluation metric. To determine
how many runs are needed, we calculated the confidence range using a 95\%
confidence interval and make sure that  the difference between the upper and
lower confidence bounds is smaller than 5\%. We instrumented the Chromium
rendering engine to measure the load time. We excluded the time spent on
browser bootstrap and shut down. To measure the energy consumption, we have
developed a lightweight runtime to take readings from the on-board energy
sensors at a frequency of 10 samples per second. We then matched the
 readings against the rendering process' timestamps to
calculate the energy consumption.

\begin{figure}[t!]
\centering
\includegraphics[width=0.4\textwidth]{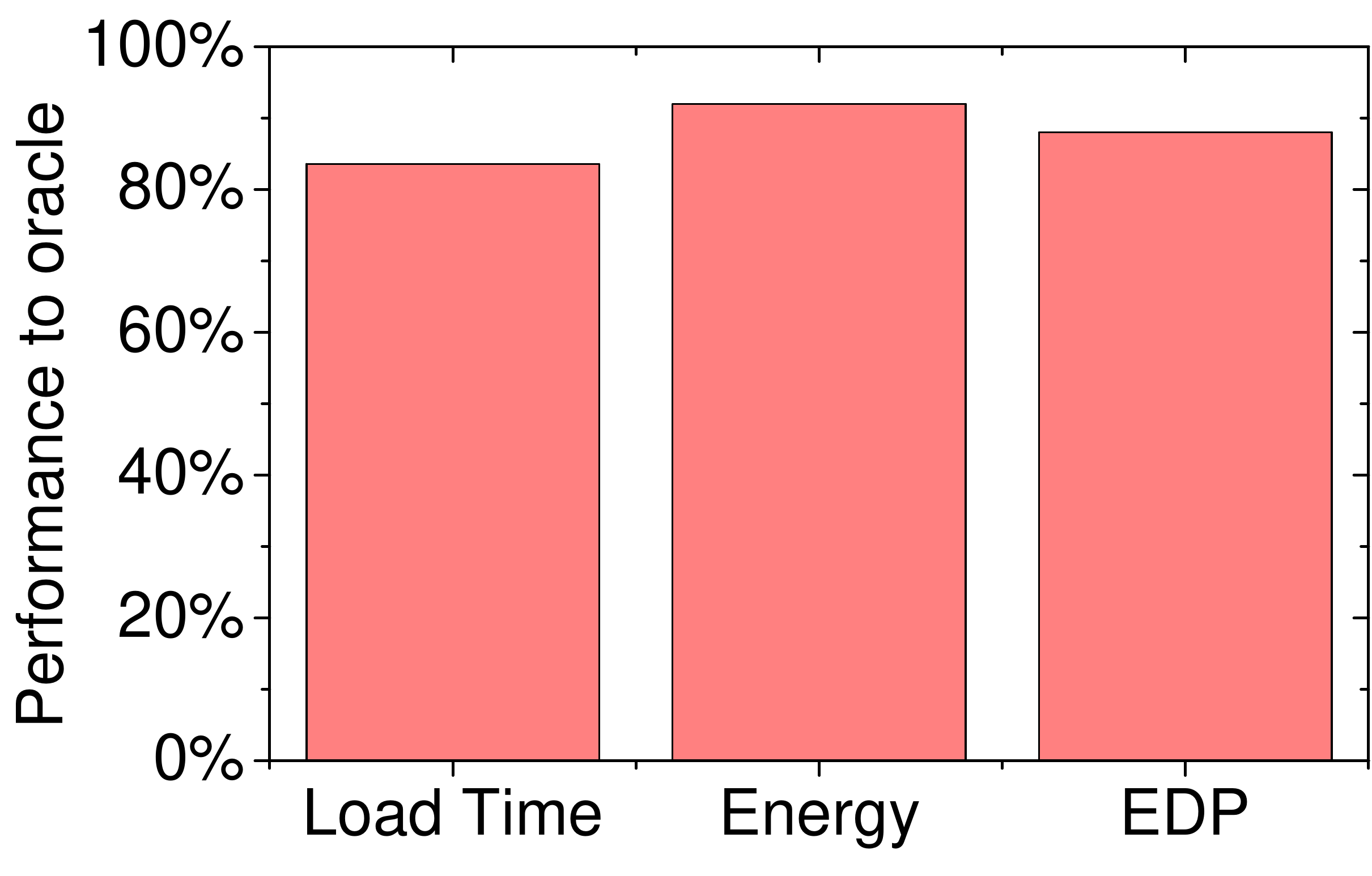}
\caption{Our performance w.r.t. performance of an oracle predictor. We
achieve over 80\% of the oracle performance.}
\label{fig:performance2oracle}
\vspace{-3mm}
\end{figure}

\section{Experimental Results}
In this section, we first compare our approach against \WS and the \HMP scheduler. We then evaluate our approach against an ideal predictor,
showing that our approach delivers over 80\% of the oracle performance.
Finally, we analyse the working mechanism of our
approach.

\subsection{Overall Results\label{sec:overall_result}}

Figure~\ref{fig:overall} shows the performance results of our approach and \WS for the three evaluation metrics across
all websites. For each metric, the performance improvement varies for different webpages. Hence, the \emph{min-max}
bars in this graph show the range of improvement achieved across the webpages we used. The baseline in the diagram is
the \HMP scheduler.

\paragraph*{Load Time} Figure~\ref{fig:overall} (a) shows the achieved performance when fast response time is the first
priority. For this metric, \WS achieves an averaged speedup of 1.34x but it causes significant slowdown (up to 1.26x)
for some websites. By contrast, our approach never leads to deteriorative performance with up to 1.92x speedup.
Overall, our approach outperforms \WS with an average speedup of 1.45x vs 1.34x over the \HMP scheduler,
and has constantly better performance across websites.

\paragraph*{Energy Consumption} Figure~\ref{fig:overall} (b) compares the achieved performance when having a long
battery life is the first priority. In this scenario, adaptive schemes (\WS and our approach) can significantly reduce
the energy consumption through dynamically adjusting the processor frequency. Here, \WS is able to reduce the energy
consumption for most websites. It achieves on average an energy reduction of 57.6\% (up to 85\%). Once again, our
approach outperforms \WS with a better averaged reduction of 63.5\% (up to 93\%). More importantly, our approach uses
less energy for all testing websites compared to \HMP, while \WS sometime uses more energy than \HMP.
This is largely due to the fact that our approach can better utilise the webpage characteristics to determine the
optimal frequencies for CPU cores. In addition, for several webpages, \WS estimates the big core gives better energy
consumption, which are actual a poor choice.

\paragraph*{EDP} Figure~\ref{fig:overall} (c) shows the achieved performance for minimizing the EDP value, i.e. to
reduce the energy consumption without significantly increasing load time. Both adaptive schemes achieve improvement on
EDP when compared to \HMP. \WS delivers on average a reduction  of 69\% (up to 84\%), but it fails to
deliver improved EDP for some websites. Unlike \WS, our approach gives constantly better EDP performance with a
reduction of at least 20\%. Overall, we achieve on average 81\% reduction (up to 95\%) of
EDP, which translates to 38\% improvement over \WS on average.

%

\subsection{Compare to Oracle}

In Figure~\ref{fig:performance2oracle}, we compare our scheme to an ideal predictor
(\emph{oracle}) that always gives the optimal processor configuration. This
comparison indicates how close our approach is to the theoretically perfect
solution.
As can be seen from the diagram, our approach achieves 85\%, 90\% and 88\% of
the oracle performance for load time, energy consumption and EDP
respectively. The performance of our approach can be further improved by
using more training webpages together with more useful features to better
characterise some of the web workloads to improve the prediction accuracy.


\subsection{Analysis}
\subsubsection{Optimal Configurations}
\begin{figure*}[t!]
	\centering
    \subfloat[Load time]
    {\includegraphics[width=0.28\textwidth]{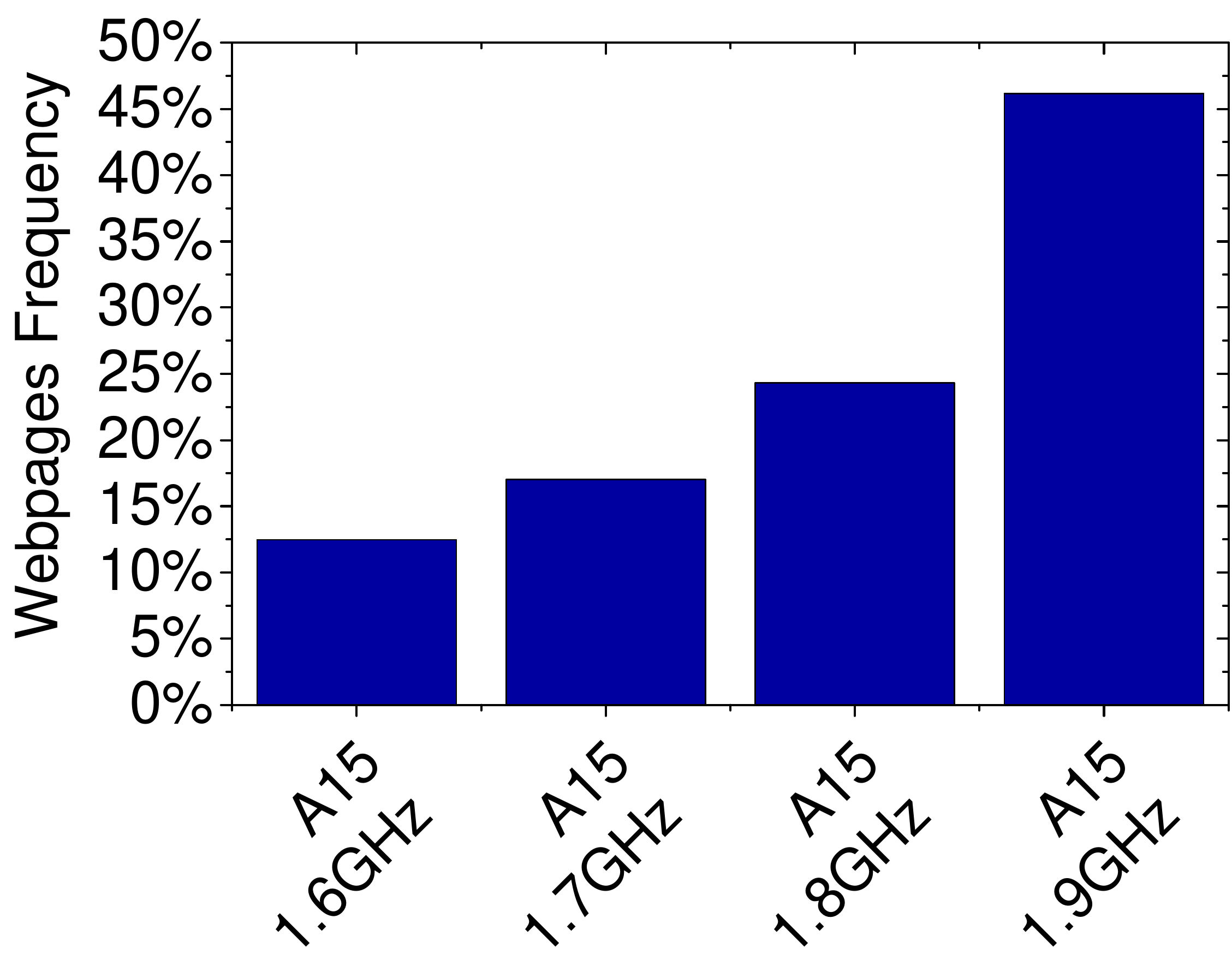}}
    \hfill
	\subfloat[Energy consumption]
    {\includegraphics[width=0.28\textwidth]{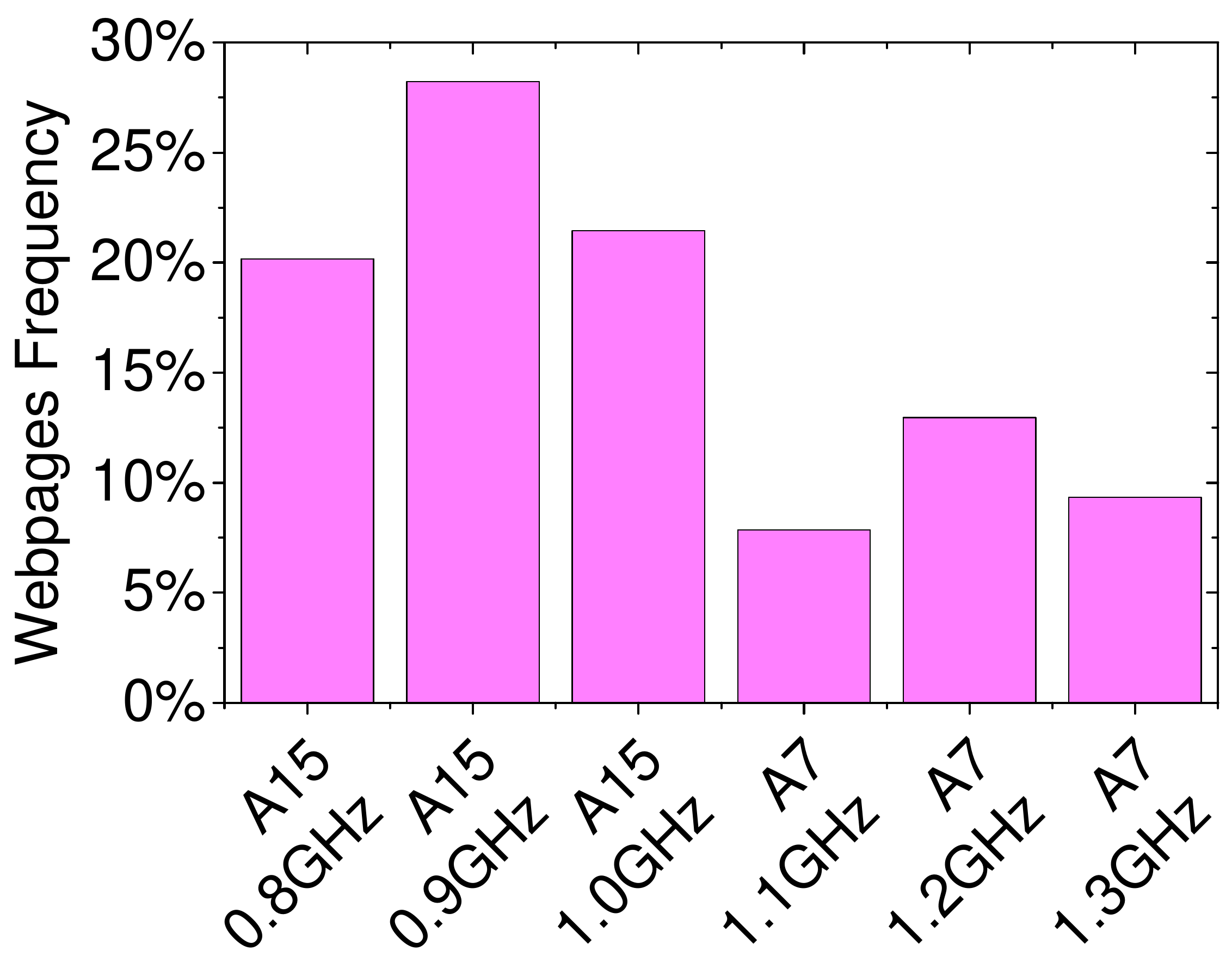}}
    \hfill
    \subfloat[EDP]
    {\includegraphics[width=0.28\textwidth]{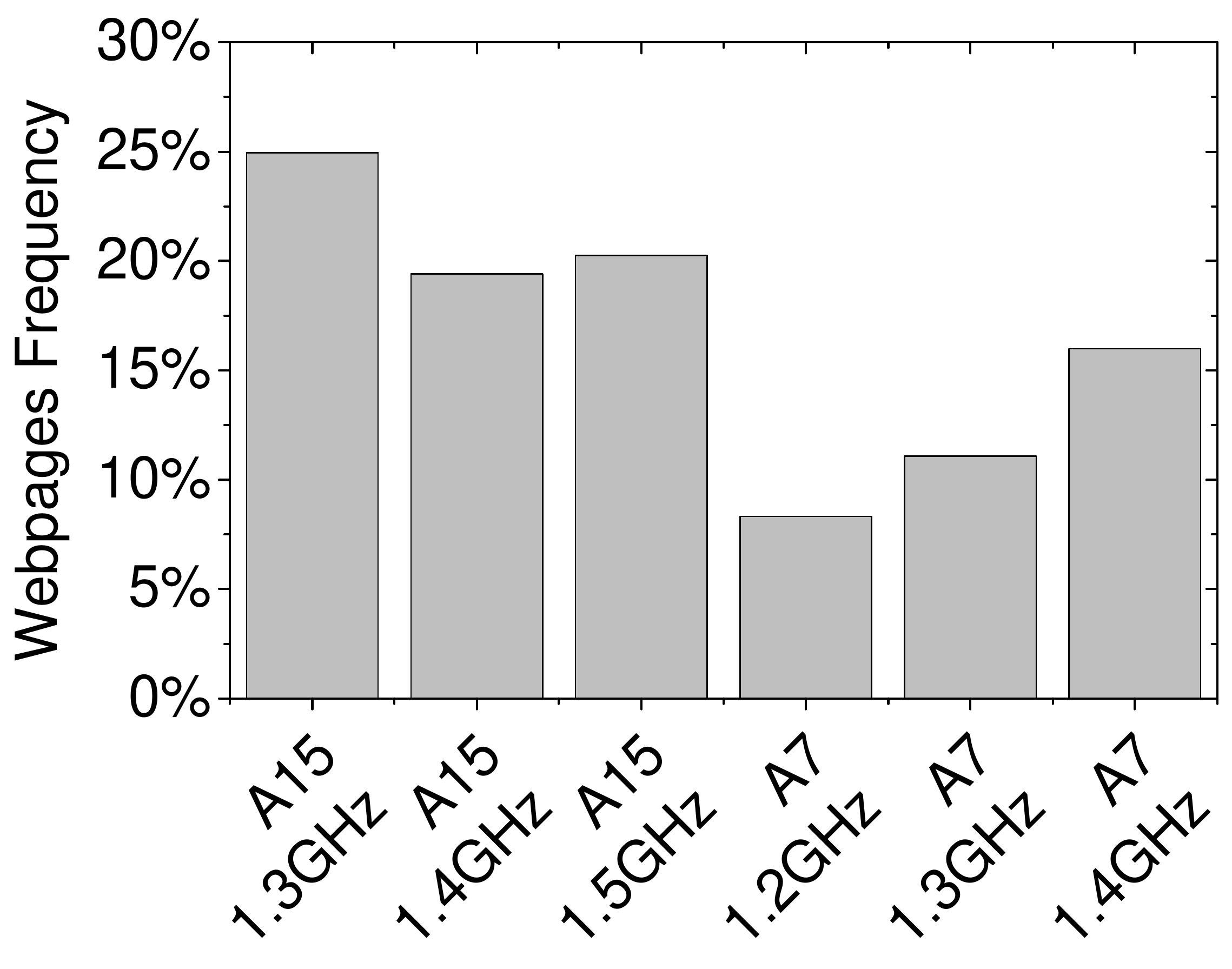}}
    \hfill
    \caption{The distribution of the optimal processor configurations for load time (a), energy consumption (b) and EDP (c).}
    \label{fig:distribution}
    \vspace{-2mm}
\end{figure*}

Figure~\ref{fig:distribution} shows how the distribution of optimal processor configurations changes from one
 metric to the other. To optimise for load time, we should always run the rendering process on the
fast, big core (A15) with a frequency of at least 1.6 GHz. For this optimisation goal, nearly half of the websites
benefit from using the A15 core at 1.9 GHz while others benefit from a lower frequency (1.6 to 1.8 GHz).
We believe this is because for some webpages using a lower frequency can reduce CPU
throttling activities~\cite{cputhrottling} (i.e. the OS will greatly clock down the processor frequency to prevent the
CPU from over-heating). We also found that running the rendering process at 2.0 GHz (a default setting used by many
performance-oriented schedulers) does not give better load time. When optimising for
energy consumption, around 30\% of the simple websites benefit from the low-power A7 core. Furthermore, for
the websites where the A15 core is a good choice, they are in favour of a lower clock frequency over the optimal
one for load time.
For EDP, using the A7 core benefits some websites but the optimal clock frequency leans towards a
median value of the available frequency range. This is expected as EDP is a metric for quantifying the trade-off
between load time and energy consumption. This diagram shows the need to adapt the
processor settings to different web workloads and optimisation goals.

\subsubsection{Performance for each configuration}
\begin{figure*}[t]
  \centering
  \includegraphics[width=0.85\textwidth]{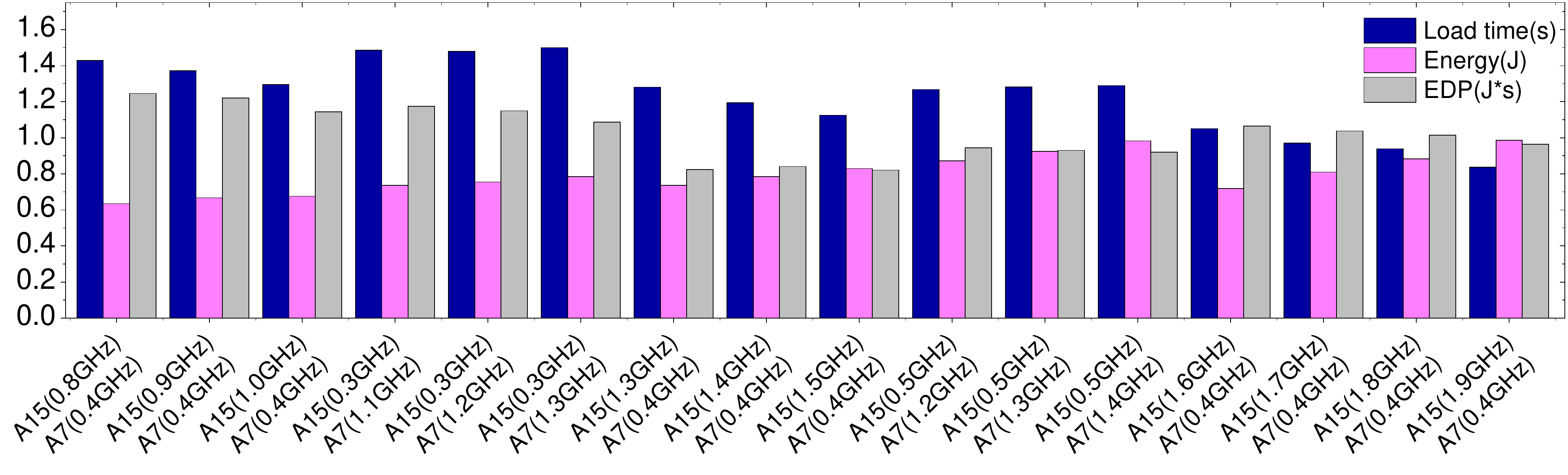}
  \vspace{-2mm}
  \caption{The achieved performance for all configurations listed in Table~\ref{tab:trainingConfig}.}
  \label{fig:all_performance}
  \vspace{-3mm}
\end{figure*}

Figure~\ref {fig:all_performance} shows the performance for using each of the processor configurations listed in
Table~\ref{tab:trainingConfig} across optimisation metrics. It shows that a ``one-size-fits-all" scheme fails to
deliver the optimal performance. For example, while the A15(0.8GHz)-A7(0.4GHz) configuration is able to reduce the
energy consumption by 40\% on average, it is outperformed by our approach that gives a reduction of 63.5\%. This is
confirmed by Figure~\ref{fig:distribution} (b), showing that running the A15 core at 0.8GHz only benefits 20\% of the
websites. Similar results can be found for the other two optimisation metrics. This experiment shows that an adaptive
scheme significantly outperforms a hardwired strategy.

\subsubsection{Prediction Accuracy}

Our approach gives correct predictions for 82.9\%, 88\% and 85\% of the webpages for load time, energy consumption and
EDP respectively. For those webpages that our approach makes a misprediction, the resulting performance is not far from
the optimal, where we still achieve a reduction of 24\%, 21\% and 56\% for load time, energy consumption and EDP when
compared to \HMP. 

\begin{figure}[t!]
\centering
\includegraphics[width=0.4\textwidth]{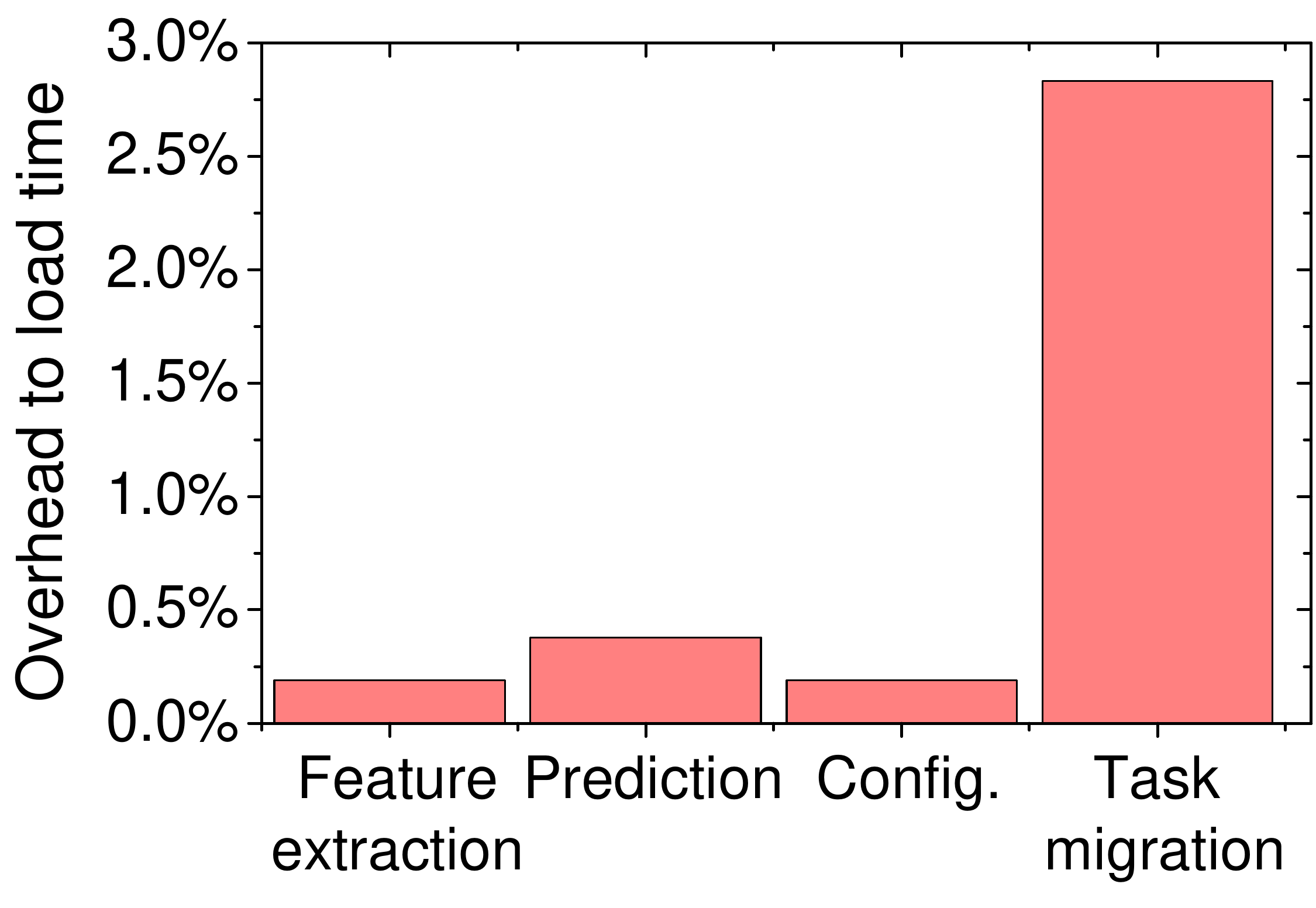}
\vspace{-2mm}
\caption{Breakdown of runtime overhead to rendering time.}
\label{fig:overhead}
\vspace{-1mm}
\end{figure}

\subsubsection{Breakdown of Overhead\label{sec:breakdown}}

Figure~\ref{fig:overhead} shows the breakdown of runtime overhead. The runtime overhead of our approach
is low -- less than 4\% with respect to the rendering time.
 Since the benefit of our approach is significant,
we believe such a small overhead is acceptable.
Most of the time (15 ms) is spent on
moving the rendering process from one processor to the other. This is expected as task migration involves initialising
the hardware context (e.g. cache warm up), which can take a few micro-seconds. The overheads of other operations, i.e.
feature extraction, predicting and frequency setting, is very low, which are less than 5
ms in total.



\subsubsection{Feature Importance}
\begin{figure}[t!]
  \centering
  \includegraphics[width=0.46\textwidth]{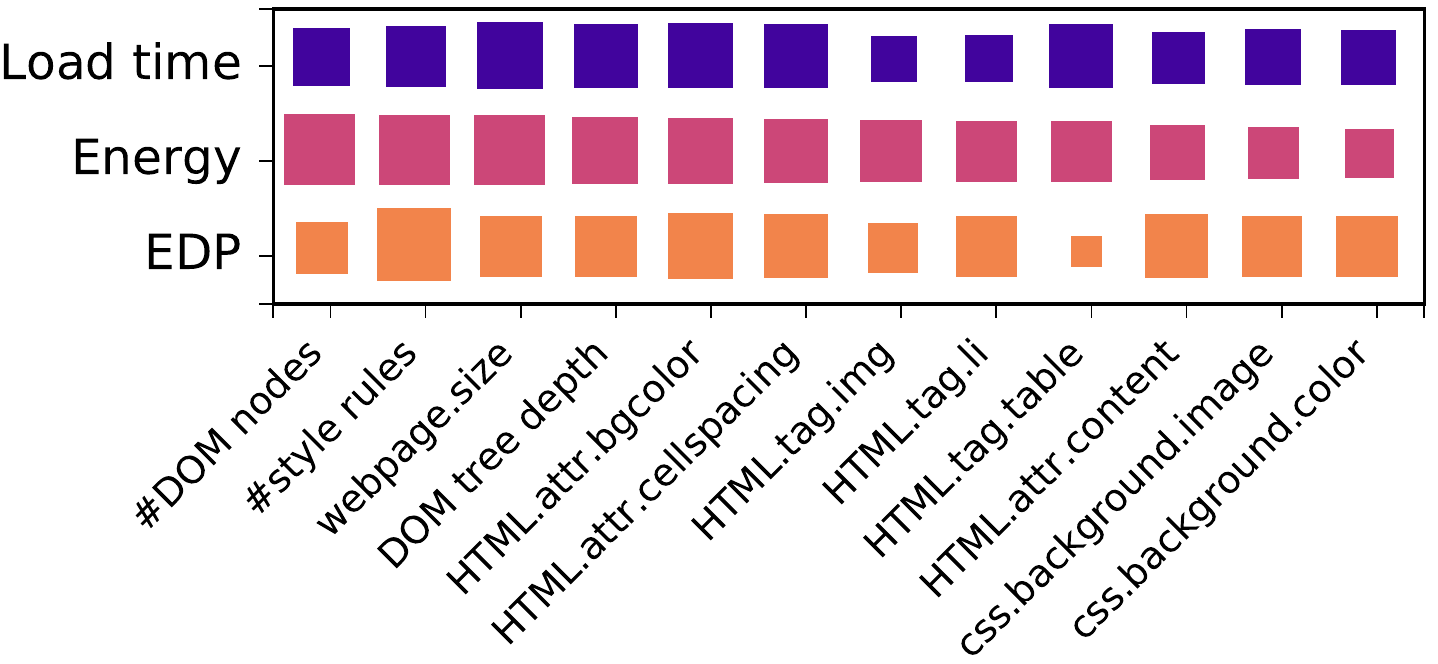}\\
  \vspace{-1mm}
  \caption{A Hinton diagram showing the importance of selected web features to the prediction accuracy. The larger the box, the more important a feature is.}
  \label{fig:hinton}
  \vspace{-4mm}
\end{figure}
Figure~\ref{fig:hinton} shows a Hinton diagram illustrates some of the most important features that have an impact on
the load time, energy and EDP specific models. Here the larger the box, the more significantly a particular feature
contributes to the prediction accuracy. The x-axis denotes the features and the y-axis denotes the model for each
metric. The importance is calculated through the information gain ratio.  It can be observed that HTML tags and
attributes (e.g. \texttt{<li>}, \texttt{<img>}, \texttt{<bgcolor>}) and style rules are important when determining the
processor configurations for all optimisation metrics. Other features are extremely important for some optimisation
metrics (such as \#DOM nodes is important for energy, and \#HTML.tag.table is important for load time and energy) but
less important for others. This diagram shows
the need for distinct models for different optimisation goals.


\subsubsection{Adapt to Different Network Environments\label{sec:network_environment}}
\begin{figure}[!t]
\begin{center}
\includegraphics[width=0.46\textwidth]{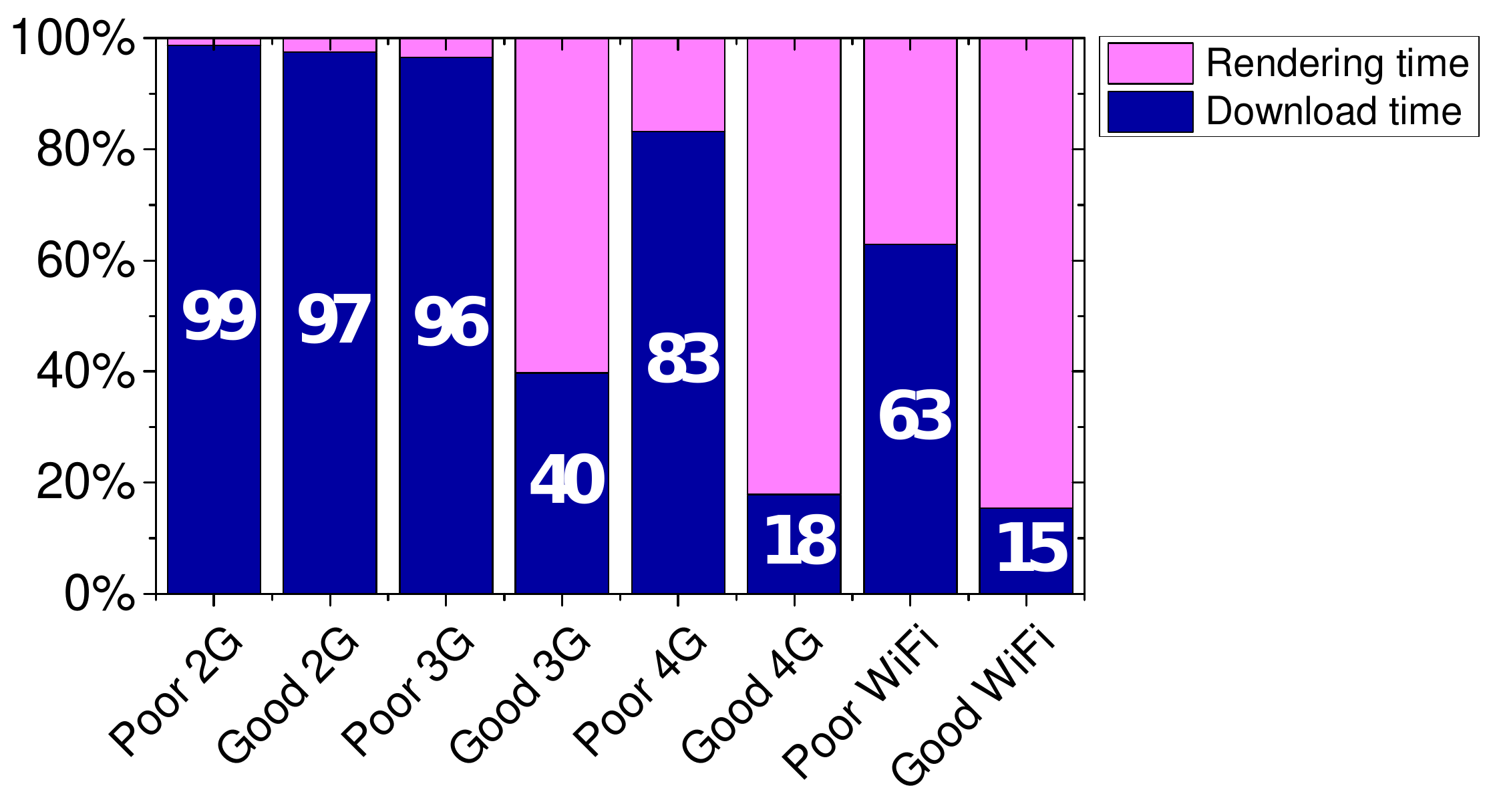}
\end{center}
\vspace{-4mm}
\caption{Webpage rendering time to the time spent on downloading the contents in different network environments.}
\vspace{-5mm}
\label{fig:latency}
\end{figure}

In all the previous experiments, we have isolated the network impact by storing the webpage into a RAM disk. In
practice, the device can be used in different network environments. A natural question to ask is: which of the three
models developed in this work best suits for a particular environment? Figure~\ref{fig:latency} shows the webpage
rendering time with respect to the download time under different network settings: 2G, 3G, 4G and WiFi (802.11). We
further breakdown each environment into two groups: poor and good. A network environment is considered to be poor if
the packet loss is greater than 30\%, otherwise it is considered to be good. As can be seen from the diagram, the
download time dominates the total processing time in poor and good 2G network environments. In such environments, our
energy-tuned model can be used to trade rendering performance for energy consumption without compromising the user
experience, by moving the rendering process to run on an low power processor at a low clock frequency. Our EDP-tuned model is
mostly suitable for a good 3G network environment with a limited download bandwidth. Finally, our load-time-tuned model
can be used in good 4G and Wifi environments to satisfy the performance requirement if load time is the first priority.
This diagram demonstrates the need of an adaptive scheme in different network environments.


\section{Related Work}
Our work lies at the intersection of numerous areas: web browsing optimisation, task scheduling, energy optimisation and predictive modeling.

\paragraph*{Web Browsing Optimisation}
A number of techniques have been proposed to optimise web browsing, through e.g. prefetching~\cite{wang2012far} and caching~\cite{qian2012web} web contents, scheduling network requests~\cite{qian2014characterizing},
or re-constructing the web browser workflow~\cite{zhao2015energy}.
Most of the prior work are built for a homogeneous  mobile systems where the processors are identical. Furthermore, prior work often targets one single optimisation goal (either performance or energy consumption).
Unlike previous research, our work targets a heterogeneous mobile system with different processing units and  multiple
optimisation goals.
The work presented by Zhu \emph{et al.}~\cite{YZhu13} is the nearest work,
which uses linear regression models to estimate the load time and energy
consumption for each web event to determine where to run the rendering
process. While promising, there are two significant shortcomings with this
approach. Firstly, it schedules the webpage to the big core with
the highest frequency if no configuration meets the cut-off
latency. This leads to poor performance as can be
seen in Section~\ref{sec:network_environment} in some networking environments.
Secondly, their linear regression
models only capture the linear correlation between the web workload
characteristics and the processor configuration, leading to a low prediction
accuracy for some webpages. Our work addresses both of these issues by
dynamically configuring all CPU cores of the system and modelling both linear
and non-linear behaviour.


\paragraph*{Task Scheduling}
There is an extensive body of work on scheduling application tasks on homogeneous and heterogeneous multi-core systems
e.g. ~\cite{zhang2002task,augonnet2011starpu,singh2013mapping}.
Most of the prior work in the area use heuristics or analytical models to
determine which processor to use to run an application task, by exploiting
the code or runtime information of the program. Our approach targets a
different domain by using the web workload characteristics to optimise mobile
web browsing for a number of optimisation metrics.


\paragraph*{Energy Optimisation}
Many techniques have been proposed to optimise web browsing at the application level. These include aggregating
data traffic~\cite{hu2014energy}, bundling HTTP requests~\cite{li2016automated}, and exploiting the radio state
mechanism~\cite{zhao2011reducing}. Our approach targets a lower level, by exploiting the heterogeneous hardware
architecture to perform energy optimisation.  Work on application-level optimisation is thus complementary to our
approach. Studies on energy behaviour of web workloads~\cite{d2016energy,thiagarajan2012killed,nicoara2015system} are also
orthogonal to our work.

\paragraph*{Predictive Modelling} Machine learning based predictive modelling
is rapidly emerging as a viable way for systems optimisation. Recent studies have shown that this technique is effective in predicting
power consumption~\cite{shye2009into}, program optimisation~\cite{1504189,zheng_pact,cgo13gwo,Wang:2014:APM:2695583.2677036,wm_lcpc,wm},
auto-parallelisaton~\cite{1542496,Wang:2014:IPP:2591460.2579561,zheng_cc14}, task
scheduling~\cite{Grewe:2011:WMA:1944862.1944881,gdwzlcpc13,EmaniW-CGO-13,7116910,middleware17}, benchmark generation~\cite{cm}, estimating
the quality of service~\cite{berral2011adaptive}, configuring processors using DVFS~\cite{kan2012eclass}, and grouping communication
traffics to reduce power consumption~\cite{tang2015energy}. No work so far has used machine learning to predict the optimal processor
configuration for mobile web browsing across optimisation goals. This work is the first to do so.


\section{Conclusions}
This paper has presented an automatic approach to optimise mobile web
browsing on heterogeneous mobile platforms, providing a significant
performance improvement over state-of-the-art. At the heart
of our approach is a machine learning based model that provides an accurate
prediction of the optimal processor configuration to use to run the web
browser rendering process, taking into account the web workload
characteristics and the optimisation goal. Our approach is implemented as an
extension to the Google Chromium web browser and evaluated on an ARM big.LITTLE mobile platform for three distinct metrics.
Experiments performed on the 500 hottest websites show that our approach achieves over 80\% of the oracle performance.
It
achieves over 40\% improvement over the Linux \HMP scheduler
across three evaluation metrics: load time, energy consumption and the energy delay product.
It
consistently outperforms a state-of-the-art webpage-aware scheduling mechanism.
Our future work will explore further refinement to prediction accuracy and to dynamically adapt to different networking environments.

\section*{Acknowledgments}
This work was performed while Jie Ren was a visiting PhD student with the School of Computing and Communications at Lancaster University.
The research was partly supported by the UK Engineering and Physical Sciences Research Council under grants EP/M01567X/1 (SANDeRs) and
EP/M015793/1 (DIVIDEND), and the National Natural Science Foundation of China under grant agreements 61373176 and 61572401. The
corresponding author of this article is Zheng Wang (email: z.wang@lancaster.ac.uk).

\bibliographystyle{abbrv}
\bibliography{refs}

\end{document}